\documentclass[%
reprint,
amsmath,amssymb,
aps,
prd,
]{revtex4-2}
\usepackage{graphicx}
\usepackage{dcolumn}
\usepackage{bm}
\usepackage{hyperref}
\usepackage[mathlines]{lineno}
\usepackage{color}

\begin{document}

	\title{Magnetorotational neutron star kicks}

	\author{Ilya A. Kondratyev}
	\email{mrkondratyev95@gmail.com}
	\affiliation{Space Reseach Institute RAS, Profsoyuznaya street 84/32, Moscow, Russia, 117997}
	
	\author{Sergey G. Moiseenko}
	\email{moiseenko@iki.rssi.ru}
	\affiliation{Space Reseach Institute RAS, Profsoyuznaya street 84/32, Moscow, Russia, 117997}
	\affiliation{National Research University Higher School of Economics, Dept. of Physics, Staraya Basmannaya street 21/4 b.5, Moscow, Russia, 105066}
	
	\author{Gennady S. Bisnovatyi-Kogan}
	\email{gkogan@iki.rssi.ru}
	\affiliation{Space Reseach Institute RAS, Profsoyuznaya street 84/32, Moscow, Russia, 117997}
	\affiliation{National Reseach Nuclear University MEPhI, Kashirskoe highway, 31, Moscow, Russia, 115409}
	

\begin{abstract}
		
   High velocity neutron stars, observed as rapidly moving radio-pulsars, are believed to gain high linear velocities -- kicks -- in aspherical supernova explosions. The mechanism of the kick formation is probably connected with anisotropic neutrino flash, and/or anisotropic matter ejection. In this paper, we investigate a neutron star kick origin in a magnetorotational (MR) supernova explosion model. The simulations have been done for a series of core collapse supernova models with initial equatorially asymmetric magnetic fields. We have realized 2D magnetohydrodynamic simulations, considering the protoneutron star kick and explosion properties in three different asymmetric magnetic field configurations. The simulations show, that in the MR supernova model protoneutron star kicks are formed with velocities up to $\sim 500$ km/s, due to asymmetric matter ejection in jets. \color{black} It may explain the observed kick velocities of some neutron stars, formed in the MR supernovae explosions. \color{black}

\end{abstract}

\maketitle

\section{Introduction}

Neutron stars and stellar mass black holes are born in explosions of massive stars known as core collapse supernovae. They are formed when the core of a massive star contracts under its own gravity. A huge amount of energy -- approximately $10^{53}$ ergs -- is released mainly in the form of neutrinos, while the energy with typical value $\sim 10^{51}$ ergs is released in the form of a supernova blast wave, enriching the Universe with heavy elements \cite{bethe1990}. After the collapse stage, a protoneutron star (PNS) or black hole is formed. During the explosion phase, due to possible non-sphericity of the explosion, a large bulk momentum -- $"$kick$"$ -- may be transferred to the PNS. The kick velocities of order of several hundreds kilometres per second are observed in a large number of pulsars \cite{lr1994, smirnova1996,  hp1997, manch2005}. A variety of different mechanisms is proposed for a natal kick gain, namely, the anisotropic supernova explosion due to global hydrodynamic instabilities \cite{kick2006, kick2013, kick2023}, a Standing Accretion Shock Instability (SASI, \cite{sasi2003}), or anisotropic neutrino flash (see, e.g., \cite{lai2001}). In the presence of magnetic field, additional anisotropy due to magnetohydrodynamic (MHD) effects is possible. The magnetic field can also start the processes of the momentum transfer from neutrinos to matter, that may occur in strong magnetic fields, which are proposed to gain additional momentum to the PNS \cite{lai2001,BK1993,AL1999,DO2020}.

The core collapse supernova explosions mechanisms are complex and involve many physical processes. Over the past 60 years, several mechanisms for the explosion of supernovae with collapsing cores have been proposed (see, for example, \cite{janka2012} for a review). Among the most developed branches of modern theory of collapsing supernovae, we can distinguish the neutrino-driven and magnetorotational (MR) mechanisms. In the framework of the neutrino mechanism \cite{colgate1966,bethe1985}, the bounce shock wave, which occurs near the protoneutron star and moves outward, is stopped by matter accreting onto the centre from the periphery of the core after passing through about $\sim 100$ kilometres. Subsequent interactions of neutrinos with matter behind the shock wave, as well as enhancing their efficiency hydrodynamic effects, such as convection \cite{epstein1979, bethe1985} and/or SASI, can lead to mixing and heating of the matter and further shock wave acceleration. Despite substantial progress in the development of the theory of the neutrino-driven supernovae, the usage of sophisticated numerical tools with multi-group neutrino transfer (see, for example, \cite{skinner2019,mueller2020}), the predicted explosion energies within the framework of the neutrino mechanism are currently significantly lower, than the observed values \cite{mueller2015}.

The MR mechanism, proposed in the work \cite{BK1970} (see also \cite{lw1970} in the context of energy release from rapidly rotating magnetized cores), on the other hand, somewhat relaxes the requirements for neutrino physics accounting, but requires the presence of rotation and magnetic field in the progenitor's core. Within the MR mechanism, after the collapse, the system represents differentially rotating protoneutron star and envelope. The frozen-in magnetic field in the system is amplified by the differential rotation, $"$winding up$"$ with time. An increase of the magnetic pressure leads to formation of a compression wave, which transforms into MHD shock wave, generating a supernova explosion. One-dimensional MHD models with simple approaches for neutrino physics \cite{bk1976,mh1979} as well as modern multidimensional calculations with various treatment of neutrinos, rotation in 2D and 3D modelling \cite{takiwaki2004, mrsn2005,mrsn2006, burrows2007, dessart2008, mueller2023} allow to obtain observed values of the explosion energy. Similar results have been obtained, using a relativistic MHD \cite{takiwaki2011, moesta2014, obergaulinger2017, aloy2021, bugli2020, kuroda2020, bugli2021, kuroda2021}. A development of magnetorotational instability (MRI, \cite{tayler1973, bh1991, akiyama2003, mrsn2005, sawai2014}) leads to exponential growth of the magnetic field in the protoneutron star, increasing the efficiency of conversion of the rotational energy into the energy of explosion \cite{akiyama2003, moesta2015}.

The MR supernovae are aspherical due to presence of rotation and aspherical magnetic fields. If the mirror symmetry of the progenitor magnetic field is violated, additional anisotropy of the outflow is formed due to different magnetic structures in southern and northern hemispheres. As a result, the magnetically driven outflows of different intensities can be formed in different hemispheres. Thus, the momentum is transferred to the PNS. The asymmetry of outflow in the framework of MR mechanism can be generated either when the poloidal magnetic field has different values in different hemispheres \cite{wang1992, sawai2008}), or by superposition of toroidal and poloidal magnetic fields of different symmetries in presence of a differential rotation \cite{BKM1992}. 
	
This paper is organized as follows. In the next section, we summarize the possible scenarios of explosion asymmetry generation, leading to PNS kick formation in the framework of MR mechanism in 2D. In the third section we discuss the initial models, while in the fourth one we report on the included physics in our MHD simulations. In the fifth section we report the results of the simulations, including  dependence of MR explosion dynamics and PNS kick and its properties on the initially asymmetric magnetic field configurations (dipolar+quadrupolar, offset dipolar and dipolar+toroidal ones) for a 35 $M_\odot$ progenitor star. The numerical technique for the simulation of MHD equations and some code tests are presented in Appendix.

\section{Different scenarios of asymmetry generation in magnetically driven supernovae}

In 2D case, the mirror symmetry violation can be of three types, which are examined in this paper. The first type is a presence of two or more multipoles with different symmetry properties with respect to equatorial plane (odd + even ones in the meaning of multipole series expansion, e.g. dipole + quadrupole, \cite{wang1992}) in the progenitor magnetic field. The second variant is an offset symmetric field configuration (see \cite{sawai2008}, where an offset strong dipole field is considered in the progenitor core in the context of rapidly moving magnetar formation). While it is unclear, how can the offset dipole field form in the progenitor core, there are some observational evidences, that such configurations may occur inside the compact objects (see \cite{wdoffset} for magnetic white dwarfs and, e.g., \cite{nsoffset} for pulsars). The presence of the offset magnetic dipole in a neutron star can also be the reason, why some pulsars are active beyond the $"$death valley$"$ \cite{arons2000}.

In presence of \color{black} the poloidal magnetic field, \color{black} the toroidal field is formed by winding in differential rotation, estimated as (see, e.g., \cite{mrsn2005,burrows2007}) \color{black}
	\begin{equation}
	\frac{\partial B_\phi}{\partial t} \sim B_p\frac{\partial \Omega}{\partial \log(r)},
	\label{winding}
	\end{equation}   
where $B_\phi, B_p$ are toroidal and poloidal magnetic field components, respectively, and $\Omega$ is an angular velocity. For pure dipole or quadrupole field the generated toroidal field will be either antisymmetric or symmetric with respect to the equatorial plane. In this case the magnetic pressure will be the same in both hemispheres, generating the symmetric explosion. In the case of purely dipole field, the toroidal field will have two extrema below and above the equator, while for the quadrupole field symmetry  the toroidal field will have a maximum value on the equatorial plane. In the case of asymmetric field, the toroidal field is amplified differently in both hemispheres, and asymmetric magnetic pressure gradient will drive the explosion differently in two hemispheres. The same stands for the offset of symmetric magnetic fields, when the offset distance from the presupernova centre is comparable with the size of the core. 
	
The third option is a presence in a presupernova core of both magnetic fields with odd (even) poloidal field and even (odd) toroidal field \cite{BKM1992}. A composition of dipole and symmetric toroidal fields is considered in this work. In presence of initial toroidal field inside the progenitor, the resulting toroidal field component will be the sum of the winded antisymmetric field, and the compressed symmetric one
 	\begin{equation}
	B_{\phi} = B_{\phi,wind}^{asym} + B_{\phi,compr}^{sym},
	\label{Btor}
	\end{equation} 
leading to magnetic pressure asymmetry with respect to the equatorial plane. 
 
During the collapse phase, the symmetric component of the toroidal field, at the magnetic flux conservation, is changing as $B_\phi^{sym} \sim \rho r$ \cite{bk1976,mh1979}. For characteristic values 
    \begin{eqnarray}
        \rho_{core} = 10^9\, {\mbox{g cm$^{-3}$}}, \,\,\,	\rho_{PNS} = 10^{14}{\mbox{g cm$^{-3}$}},\,\,\,
        \nonumber\\
        r_{core}=1000\, {\mbox{km}},\,\,\,
        r_{PNS}=10\, {\mbox{km}},\,\,\,
        B_{\phi,core}^{sym} = 10^{13}\, {\mbox{G}} \qquad
        \label{par}
    \end{eqnarray} 
we obtain formation of the PNS with $B_{\phi,PNS}^{sym} =B_{\phi,core}^{sym} (\rho_{PNS}\, r_{PNS}/ \rho_{core}\, r_{core})= 10^{16}$ G. In our calculations this value coincides with a characteristic value reached by $B_{\phi,PNS}^{asym}$ during its formation in the collapse. Therefore at the parameters from \eqref{par} the initial field $B_{\phi,core}^{sym}$ effectively influences on the explosion process, initiating the asymmetry of jets.

\section{Initial models}

In this work, we simulate the magnetically driven supernova explosion with asymmetric magnetic fields. We utilize a progenitor model \color{black} $35OC$ \color{black} of a massive 35 $M_\odot$ zero-age main sequence rotating magnetized star from \cite{woosley2006}, which was used in a number of papers, devoted to MR explosions (see, e.g., \cite{dessart2008, obergaulinger2017, bugli2020, varma2021} in a 2D framework). The iron core radius of this star is $2890$ kilometres at the onset of collapse, \color{black} and it has a mass $M_{core}\approx 2.1 M_\odot$. \color{black} For this study we use a progenitor with an angular velocity depending on the spherical radius \cite{woosley2006}, with the ratio of rotational to gravitational energies $T/|W| \eqsim 0.14\%$. 
\color{black} The angular velocity of the presupernova model is $\Omega \approx 2$ rad/s at the center of the star, while it is approximately an order of magnitude smaller at the periphery of the core. 

\color{black} 

We consider three different variants -- superposition of dipole and quadrupole fields \cite{bugli2020}, an offset dipole field, and a superposition of a dipole and a toroidal field, which is even with respect to the equatorial plane \cite{suwa2007, mueller2023}. Vector potentials for these configurations can be written as 
	\begin{equation}
	\begin{gathered}
	A_\phi^{dip} = r\frac{B_{0,dip}}{2}\frac{r_0^3}{(r^2 - 2r\cos\theta z_{off} + z_{off}^2)^{3/2}+r_0^3}\sin \theta, \\
	A_\phi^{quad} = r\frac{B_{0,quad}}{2}\frac{r_0^4}{r^4+r_0^4}{\cos \theta \sin \theta}, \\
	A_r^{tor} = r\frac{B_{0,tor}}{2}\frac{r_0^3}{r^3+r_0^3}\cos\theta.
	\end{gathered}
	\label{magfield}
	\end{equation}
The vector potentials $A_\phi^{dip}$ and $A_\phi^{quad}$ set up the poloidal magnetic field of dipolar and quadrupolar types, respectively, while $A_r^{tor}$ sets the toroidal one. Parameters $B_{0,*}$ (where symbol $*$ stands for $dip,quad,tor$) and $r_0$ allow us to regulate the field strength and its localization degree in the core, respectively. Within the region with $r\ll r_0$ (for an offset parameter $z_{off}=0$) these fields are close to the uniform ones with the induction value $B \sim B_0$, while for $r \gg r_0$ they tend to decrease according to the formulae in vacuum. The magnetic field is generated by electrical currents inside the radius $r \lesssim r_0$, with different types of the symmetry. Parameter $z_{off}$ in dipolar potential sets the origin of the field, which is offset from the centre at the distance $z_{off}$ along the rotational axis.

	\begin{table}[h!]
		\begin{center}
			\caption{ The models with different magnetic fields in the progenitor core. $"DQ"$ models are related to dipolar + quadrupolar superposition with zero offset; $"Do"$ models correspond to offset dipolar magnetic fields; $"DT"$ models correspond to a superposition of dipolar and toroidal fields.  See the text for more details. }
			\label{tab:table1_models}
			\begin{tabular}{c|c|c}
				
				\textbf{model} & \textbf{parameters} & \textbf{comment} \\
				
				\hline
    
				$DQ-1e10$ &  $B_{0,dip; quad} = 1\cdot10^{10}$ G   & dip + quad \\
				$DQ-2e10$ &  $B_{0,dip; quad} = 2\cdot10^{10}$ G   & dip + quad \\
				$DQ-5e10$ &  $B_{0,dip; quad} = 5\cdot10^{10}$ G   & dip + quad \\
				$DQ-1e11$ &  $B_{0,dip; quad} = 1\cdot10^{11}$ G   & dip + quad \\
				$DQ-3e11$ &  $B_{0,dip; quad} = 3\cdot10^{11}$ G   & dip + quad \\
				$DQ-6e11$ &  $B_{0,dip; quad} = 6\cdot10^{11}$ G   & dip + quad \\
				$DQ-1e12$ &  $B_{0,dip; quad} = 1\cdot10^{12}$ G   & dip + quad \\
				
				\hline
    
				$Do-3e10$ & $B_{0,dip} = 3\cdot10^{10}$ G &  offset dip, $z_{off} = 1500$ km \\
				$Do-2e11$ & $B_{0,dip} = 2\cdot10^{11}$ G &  offset dip, $z_{off} = 1500$ km \\
				$Do-1e12$ & $B_{0,dip} = 1\cdot10^{12}$ G &  offset dip, $z_{off} = 1500$ km \\
				
				\hline
				
				$DT-4e10$ & $B_{0,dip} = 4\cdot10^{10}$ G & dip + tor, $B_{0,tor} = 10^{13}$ G \\
				$DT-6e10$ & $B_{0,dip} = 6\cdot10^{10}$ G & dip + tor, $B_{0,tor} = 10^{13}$ G \\
				$DT-1e11$ & $B_{0,dip} = 1\cdot10^{11}$ G & dip + tor, $B_{0,tor} = 10^{13}$ G \\
				$DT-3e11$ & $B_{0,dip} = 3\cdot10^{11}$ G & dip + tor, $B_{0,tor} = 10^{13}$ G \\
				$DT-6e11$ & $B_{0,dip} = 6\cdot10^{11}$ G & dip + tor, $B_{0,tor} = 10^{13}$ G \\
				$DT-1e12$ & $B_{0,dip} = 1\cdot10^{12}$ G & dip + tor, $B_{0,tor} = 10^{13}$ G \\
				
			\end{tabular}
		\end{center}
	\end{table}

We simulated here 16 non-symmetric models, which are listed in the Table \ref{tab:table1_models}. In the Table, $"DQ"$ models are related to dipolar + quadrupolar superposition, where both components have the same strength. Parameter $r_0 = 2000$ km is chosen for all $"DQ"$ simulations with $z_{off} = 0$. In presence of the quadrupolar field, we normalize the progenitor magnetic fields so, that the total magnetic energy inside the sphere of radius $r_0$ is equal to the energy of centred dipole with the field strength $B_{0,dip}$. $"Do"$ models correspond to offset dipolar magnetic fields.  We impose $z_{off} = r_{0} = 1500$ km in order to consider configurations with well pronounced influence of the offset parameter. In $DQ$ and $Do$ models the initial magnetic field is localized mostly in the northern hemisphere, where both components have the same directions of the field, and in the offset the center is always moved upstairs. The $"DT"$ models with initial toroidal fields have large toroidal component in the progenitor $B_{0,tor} = 10^{13}$ G. For all of them we set $r_0 = 2000$ km and $z_{off} = 0$.

\section{Input physics}

We have solved non-relativistic magnetohydrodynamic (MHD) equations for series of models, described in the previous section. In the numerical modeling we used 2D, axially symmetric $(\partial/\partial\phi=0)$ MHD equations in spherical coordinates $(r,\, \theta,\, \phi)$, with approximate neutrino cooling/heating source terms and self-gravity. The following equations have been solved:

\begin{widetext}
	\begin{equation}
	\begin{gathered}
	\partial_t \rho + {\rm div} ( \rho\textit{\textbf{v}}) = 0, \\
	\partial_t(\rho v_r) + {\rm div} \left( \rho v_r \textit{\textbf{v}} - \frac{B_r\textit{\textbf{B}}}{4\pi}\right) + \frac{\partial P_{tot}}{\partial r} = \rho\frac{v_\theta^2+v_\phi^2}{r}-\frac{B_\theta^2+B_\phi^2}{4\pi r} - \rho \frac{\partial \Psi}{\partial r} \\ 
	\partial_t (\rho v_\theta) + {\rm div} \left(\rho v_\theta \textit{\textbf{v}} - \frac{B_\theta \textit{\textbf{B}}}{4\pi}\right) + \frac{1}{r}\frac{\partial P_{tot}}{\partial \theta} = \rho\frac{ v_\phi^2\cot\theta - v_rv_\theta}{r} + \frac{B_rB_\theta - B_\phi^2\cot\theta}{4\pi r} - \frac{\rho}{r}\frac{\partial \Psi}{\partial \theta} \\
	\partial_t(\rho v_\phi) + {\rm div} \left( \rho v_\phi \textit{\textbf{v}} - \frac{B_\phi\textit{\textbf{B}}}{4\pi}\right) = \frac{B_\theta B_\phi\cot\theta-B_rB_\phi}{4\pi r} - \rho\frac{ v_rv_\phi + v_\theta v_\phi\cot\theta}{r}, \\
	\partial_t E + {\rm div}\left((E+P_{tot})\textit{\textbf{v}} - \frac{\textit{\textbf{B}} (\textit{\textbf{B}} \cdot\textit{\textbf{v}})}{4\pi}\right) = -\rho\textit{\textbf{v}}\cdot{\nabla}\Psi + Q_{\nu}, \\
	\partial_t(\rho Y_e) + {\rm div} ( \rho Y_e\textit{\textbf{v}}) = R_{\nu}, \\
	\partial_t\textit{\textbf{B}} - {\nabla}\times(\textit{\textbf{v}}\times \textit{\textbf{B}}) = 0, \\
	\Delta \Psi = 4\pi G \rho.\\
	\end{gathered}
	\label{MHD}
	\end{equation}
\end{widetext}

In this system  $\rho, P, Y_e$ are the plasma density, pressure and electron fraction (the number of unpaired electrons per baryon, $Y_e = \frac{n_e^- - n_e^+}{\Sigma_{i} A_i n_i + n_p + n_n }$, where $n_e^-, n_e^+, n_i, n_p, n_n$ are number densities of electrons, positrons, nuclei, protons and neutrons, correspondingly, while $A_i$ is a nuclei mass number), $\textit{\textbf{B}}, \textit{\textbf{v}}$ are the magnetic field and velocity vectors, $E = \rho\frac{\textit{\textbf{v}}^2}{2} + \frac{\textit{\textbf{B}}^2}{8\pi} + \rho e_{int}$ is a total energy density, where $e_{int}$ is a specific internal energy. $P_{tot} = P + \frac{\textit{\textbf{B}}^2}{8\pi}$ is a total pressure. $\Psi$ is a gravitational potential. $R_\nu$ is the term, defining the change of $Y_e$ due to weak processes, $Q_\nu$ is a neutrino source term (cooling or heating) for the thermal energy changes. 
	
For all models we use Shen \cite{shen2011} equation of state (EOS) at densities $\rho>10^8$ g/cm$^{3}$, with account of different nuclei in equilibrium, electrons, positrons, and black-body radiation. For lower densities we used the EOS with electrons, positrons, ideal gas of baryons, and a black-body radiation with a fixed compositions at the border density $\rho=10^8$ g/cm$^{3}$. The resulting table was taken from \cite{oconnorott2010}. 
	
To simulate the neutrino physics, for all models we used the multi-flavour neutrino leakage scheme, following \cite{oconnorott2010} without inclusion of $\mu,\tau$-neutrinos and a pressure of trapped neutrinos, with runaway neutrino for $R_\nu$. The term $Q_\nu$ includes cooling \cite{rl2003, ruffert1996} and heating terms. Heating is important after a core bounce, and is taken from \cite{janka2001}. Before the core bounce we use the prescription of matter deleptonization developed in \cite{liebendorfer2005}. 
 
A self-gravity is simulated by the Legendre polynomials expansion method \cite{steinmetz1995} with a general relativistic correction \cite{marek2006} for the monopole term. 
  
We simulate the MHD equations, using the newly developed Godunov-type high-order explicit MHD code, the details and code tests are reported in Appendix. The numerical resolution for all runs is $N_r = 360, N_\theta = 128$ along $r$ and $\theta$ directions, respectively ($0 \le r \le 47.000$ km, $0 \le \theta \le \pi$). The radial grid for $r \le 14$ km is uniform with cell size $\Delta r = 500$ m.  For $14$ km $\le r \le 47.000$ km the grid in radial direction is rarefied as $\Delta r_i = r_{i-1}\frac{\pi}{N_\theta}$ ($i$ is a cell radial index). The grid in $\theta$-direction is equally spaced for every $r_i$ and constructed in the following way -- we start with 2 cells for $\Delta r_1$, then the number of grid points in $\theta$-direction grows with increasing of radius $r$ in order to relax a restrictive Courant-Friedrichs-Lewy condition due to converging cell size near the coordinate center (see Appendix).

\section{Explosion dynamics and kick properties}

\subsection{General considerations}\label{sect51}

The collapse proceeds essentially spherically-symmetric until the  stages, when the centrifugal force is becoming important. A bounce time is almost independent on the magnetic field strength and its configuration, and it lasts $\sim 320$ ms from the onset of the collapse. In general, all models show usual dynamics for the MR supernovae. After the collapse stage and a bounce shock formation, the toroidal field starts to amplify in all layers of the PNS and surrounding medium. A formation of a convectively unstable layer happens \cite{epstein1979} after the bounce, where the electron fraction gradient $\frac{dY_e}{dr} < 0$. \color{black} This layer is formed inside the region between $10$ km up to $\sim 50$ km from the star's center.

The presence of a turbulent motion triggered by convection in the magnetized medium of the PNS could potentially provide a dynamo field amplification mechanism,  generating very large magnetic fields inside the PNS  (see \cite{td1993} and 3D calculations \cite{raynaud2020}). In our two-dimensional setting, the dynamo action is prohibited according to the Cowling theorem. A convective flux expulsion effect takes place in the  convective layer region, see also \cite{obergaulinger2014}. It consists of decreasing of the field in the convective shell looking out as the field lines being forced out (expelled) onto the boundaries of the convective layer, see \cite{obergaulinger2014, matsumoto2020} and the right panel in \eqref{fig:bfield_dq} in the next subsection of this paper. The form of the poloidal field configuration at late post-bounce times is almost independent on the initial precollapse magnetic field. In  2D axisymmetry assumption used here, the toroidal field amplification in the differentially rotating PNS happens due to the wrapping of the field lines, according to Eqn. \eqref{winding}.  The poloidal component of the field is amplified due to a possible action of the magnetorotational instability and a magnetized matter fall-back from the exterior parts of the collapsing core.
\color{black}

At increasing of the magnetic field, the magnetic pressure gradient sets up an explosion and jet-like outflows. The onset of explosion depends significantly on the value of initial magnetic field. In the models with highest used initial poloidal field  $B_p \sim 10^{12}$ G, the explosion and jet formation start almost promptly, with explosion time $t_{expl} < 30$ ms. For low poloidal precollapse field $B_p \sim 10^{10}$G, the explosion is delayed, and the envelope starts to expand at $t_{expl} \sim 400$ ms after the core bounce. \color{black} In the simulated models, we have obtained the PNSs with high masses of order of $\sim 2.2 M_\odot$. In our calculations, a major part of the PNS mass accumulates by a matter fall-back during the first $400$ ms after the core bounce, in accordance with, e.g., \cite{obergaulinger2017}. \color{black}

We calculate the energy of explosions, using the following formula
	\begin{equation}
	\begin{gathered}
	E_{expl} = \int_{E+\rho\Psi>0}(E+\rho\Psi) dV,
	\end{gathered}
	\label{expldata}
	\end{equation}
where $E = \rho\frac{\textit{\textbf{v}}^2}{2} + \frac{\textit{\textbf{B}}^2}{8\pi} + \rho e_{therm}$, and integration is performed over the region, where the matter has a local positive energy, including the gravitational binding energy. The kick velocity value $v_{kick}$ is calculated \color{black} indirectly\color{black}, assuming momentum conservation of the whole system during the explosion 
	\begin{equation}
	v_{kick} = -\frac{1}{M_{PNS}} \int_{E+\rho\Psi>0}\rho v_z dV = -\frac{p_{eject,z}}{M_{PNS}}.
	\label{kick_vel}
	\end{equation}
The PNS mass $M_{PNS}$ is calculated as a volume integral over the region, where the density is larger than $10^{11}$ g/cm$^{3}$; $p_{eject,z}$ is a $z$-component of the momentum of the ejected matter. The determination of the kick velocity value in \eqref{kick_vel} is similar to \cite{mueller2023}.

\color{black} 
   
In this problem, the speed acquired by the PNS is much smaller than the speed of jets, which are responsible for the PNS kick motion: several hundreds km/s for the PNS relative to several tens of thousands of the jet speed \cite{kick2013}. Therefore, the PNS motion is neglected in calculations and is absent in figures.
The magnetic forces, which trigger the MR explosion, are concentrated in the vicinity of the PNS, and the system works similarly to a $"$jet engine$"$, leading to the outflows with different intensities with respect to the equatorial plane. Due to the momentum conservation, we can indirectly evaluate the velocity of the PNS based on the ejected momentum from both hemispheres according to Eqn. \eqref{kick_vel}. 

\color{black}

The level of the explosion asymmetry is represented by the value $A_E$ as follows:
	\begin{equation}
	A_E = \frac{E_N-E_S}{E_N+E_S},
	\label{an_par}
	\end{equation}
where $E_N$ and $E_S$ are parts of the energy in the ejected material from northern and southern hemispheres, respectively. It lies in the interval of a range $(-1,1)$. Positive values of $A_E$ correspond to the case, where the explosion in the northern direction is stronger, while the negative ones correspond to stronger explosion in the southern hemisphere. A zero value of $A_E$ defines a symmetric explosion. The sign of the kick velocity is opposite to the sign of $A_E$.

The kinetic energy of the PNS kick is very low compared to the supernova explosion energy. \color{black} For instance, even for a massive neutron star with the mass $2 M_{\odot}$ and $v_{kick}=300$ km/s, $E_{kick} = 9 \cdot 10^{47}$ ergs, \color{black} while the explosion energy in our calculations is $E_{expl} \sim 10^{51}$ ergs. It is necessary to ensure, that  anisotropy connected with numerical errors is low enough, and does not produce an artificial asymmetry of the explosion. In order to do this, we have simulated the MR explosion with a symmetric magnetic field. We considered a model with the purely dipolar field, with parameters $B_0 = 10^{12}$ G, $r_0 = 2000$ km and $z_{off} = 0$ from \eqref{magfield}, which is referred as $"D0-1e12"$. The explosion energy in this model is $E_{expl} \simeq 1.11\cdot10^{51}$ ergs, and the kick velocity is $v_{kick} \simeq 6.46$ km/s at $661$ milliseconds after the core bounce. Hence, the level of numerical anisotropy is low enough to be certain in the results about asymmetry properties in our calculations.

\subsection{Results for the superposition of dipolar and quadrupolar fields}

\begin{figure}[!htp]
		\centering
        \includegraphics[width=8.6cm,height=6.0cm]{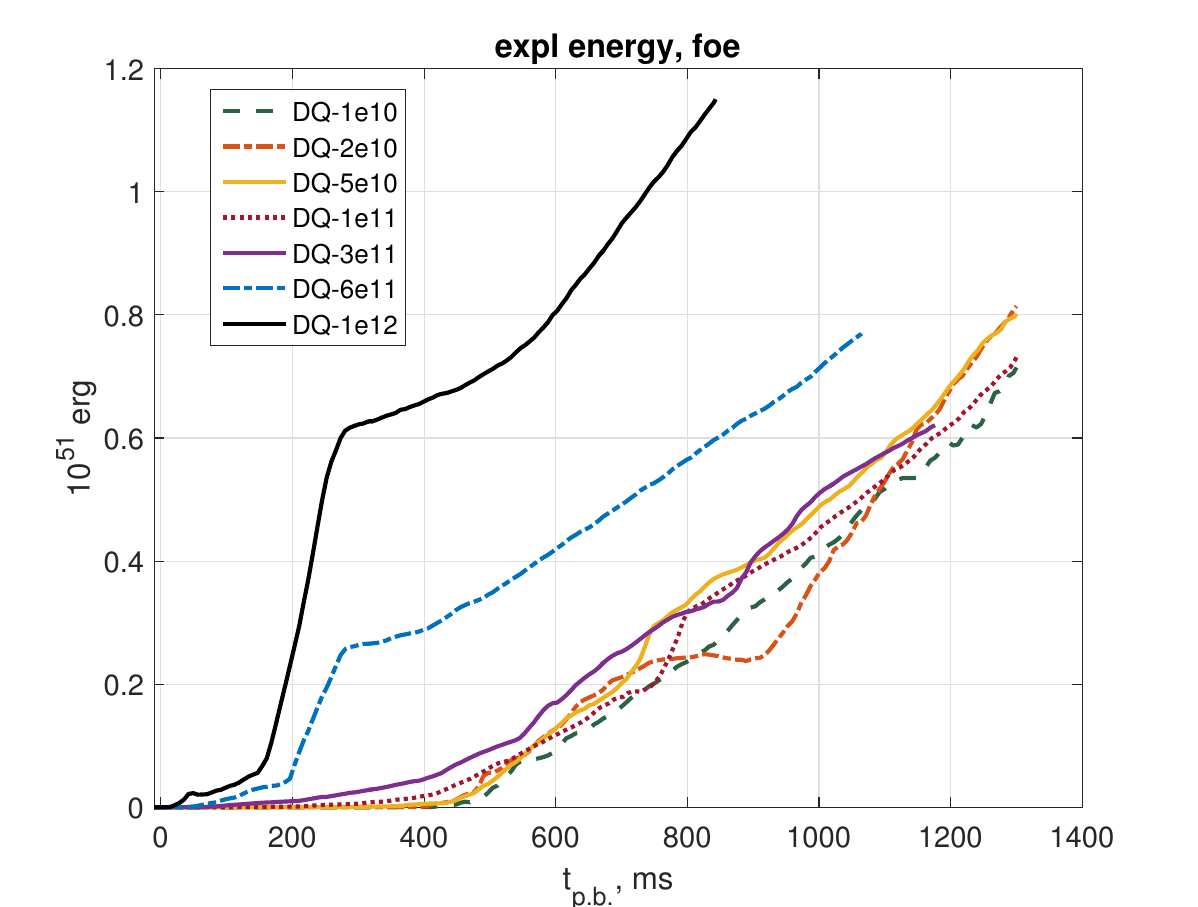}
		\includegraphics[width=8.6cm,height=6.0cm]{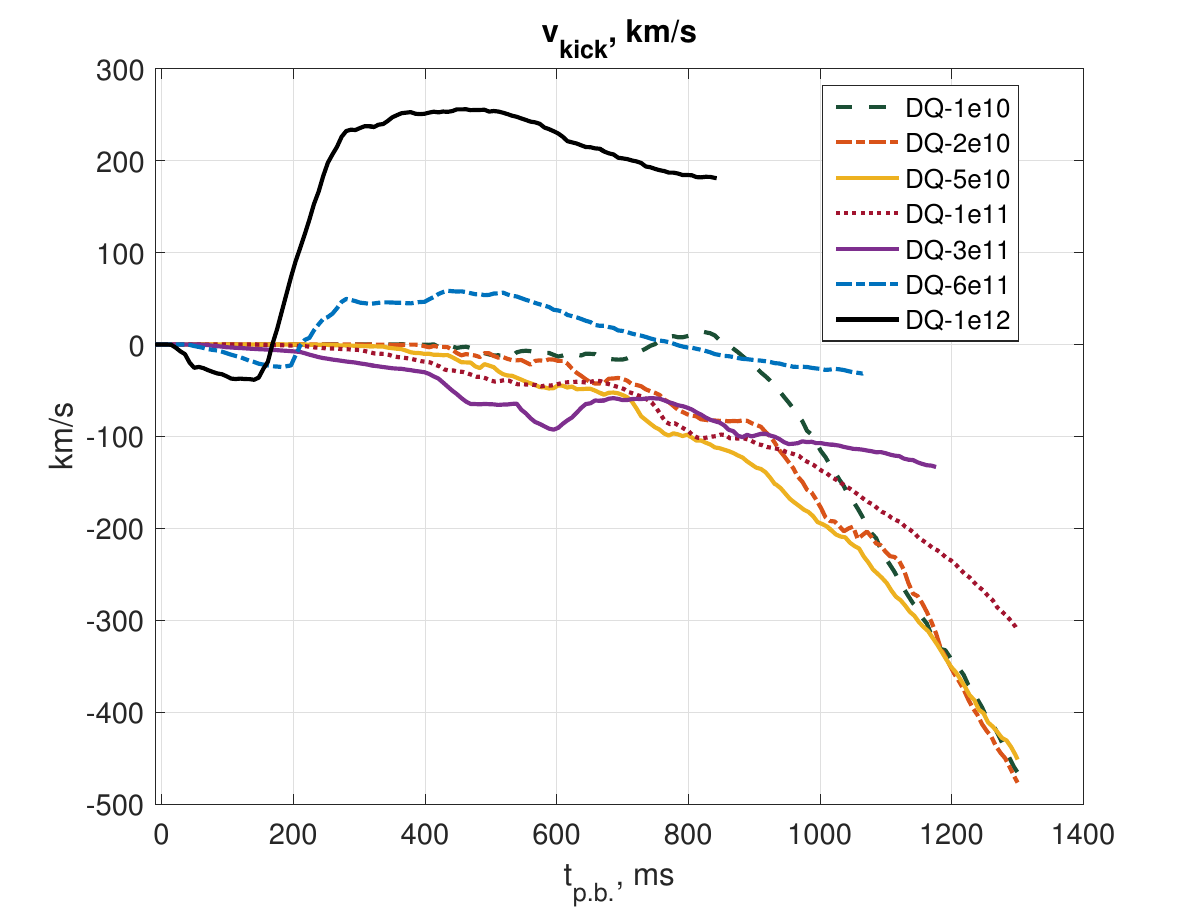}
		\caption{ Integral characteristics for configurations with dipolar + quadrupolar magnetic fields as the functions of the post-bounce time: upper panel -- explosion energy, lower panel -- PNS kick velocity. }
		\label{fig:expl_dq}
	\end{figure}

The energies of explosions, PNS kick velocities and explosion anisotropy values, obtained  for a superposition of dipolar + quadrupolar fields, are presented in Table \ref{tab:table_results_dq}. Figure \ref{fig:expl_dq} shows the dependence of explosion energies and PNS kick velocities on the post-bounce time for $DQ$ models from the Table \ref{tab:table1_models}.

	\begin{table}[h!]
		\begin{center}
			\caption{The simulation results for the models with a superposition of dipolar and quadrupolar fields. Explosion energies, PNS kick velocities and  anisotropy values $A_E$ are shown.  The final post-bounce times are given in the last column. }
			\label{tab:table_results_dq}
			\begin{tabular}{c|c|c|c|c}
				\textbf{model} & $E_{expl}$, $10^{51}$ ergs & $v_{kick}$, km/s & $A_E$ & $t_{p.b.}^f$, ms \\
				\hline
				$DQ-1e10$ & 0.715 & -465.5 & 0.793 & 1300 \\
				$DQ-2e10$ & 0.814 & -476.7 & 0.696 & 1300 \\
				$DQ-5e10$ & 0.802 & -451.6 & 0.665 & 1300 \\
				$DQ-1e11$ & 0.731 & -311.1 & 0.532 & 1300 \\
				$DQ-3e11$ & 0.621 & -133.2 & 0.206 & 1176 \\
				$DQ-6e11$ & 0.770 & -31.45 & -0.026 & 1065 \\
				$DQ-1e12$ & 1.151 &  181.0 & -0.277 & 843 \\
			\end{tabular}
		\end{center}
	\end{table}

\color{black}
In all $DQ$ models jets are forming first in the northern hemisphere, where the multipole components are summed up. However, the morphology of the blast wave evolves differently in cases of strong magnetic fields in the models with $B_0=6\cdot 10^{11}$ and $10^{12}$ G, from the ones with smaller values of $B_0$.

\begin{figure}[!htp]
		\centering
        \includegraphics[width=8.6cm,height=7.5cm]{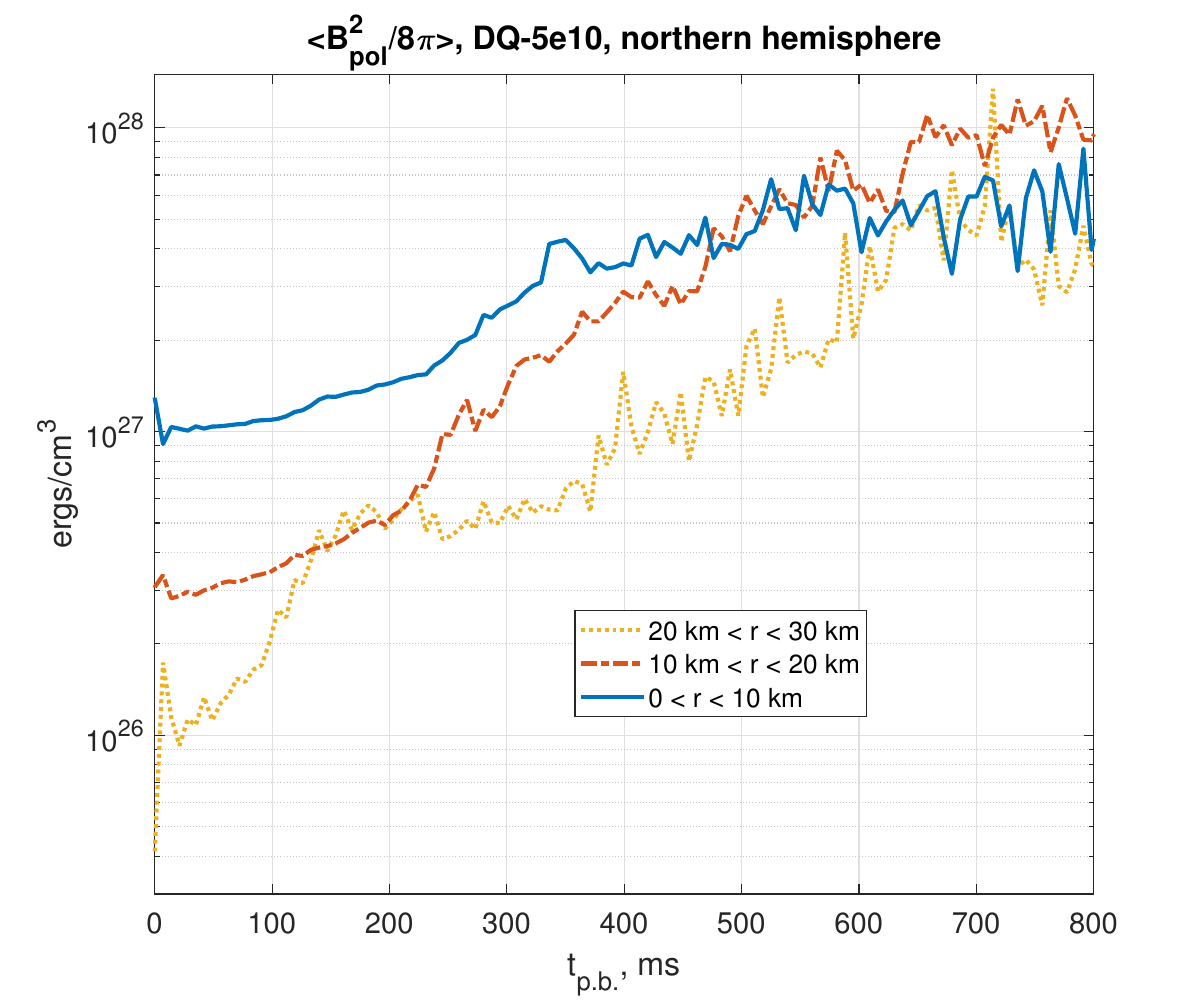}
		\includegraphics[width=8.6cm,height=7.5cm]{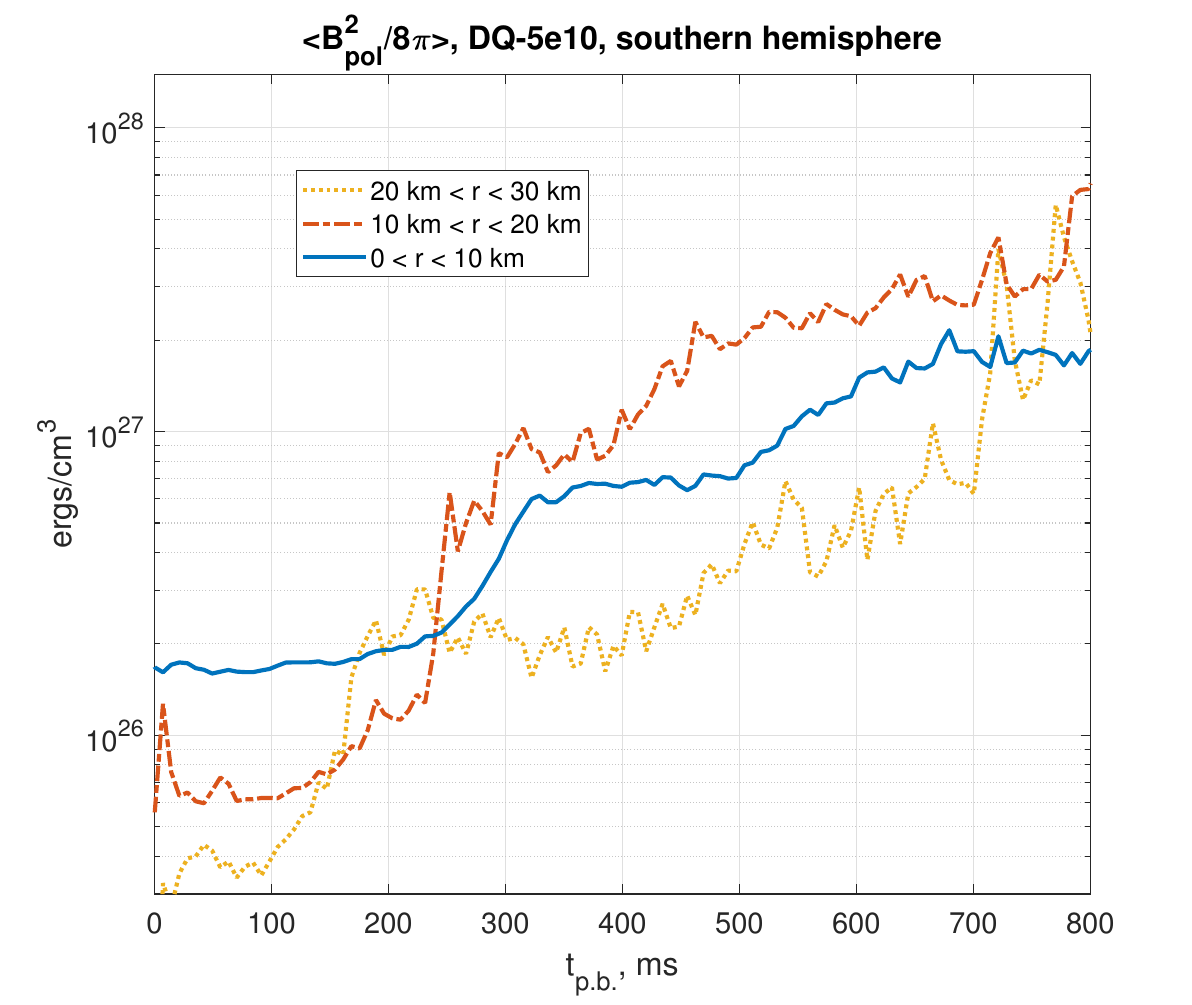}
		\caption{ Averaged density of poloidal magnetic energy in three spherical shellular regions within $0 < r < 10$ km (blue solid lines), $10$ km $ < r < 20$ km (red dash-dotted lines) and $20$ km $ < r < 30$ km (yellow dotted lines) inside the PNS as the function of the post-bounce time for the model $DQ-5e10$. Upper panel -- northern hemisphere, lower panel -- southern hemisphere. }
		\label{fig:dq5e10_b_evol}
	\end{figure}

We see in Fig.\ref{fig:expl_dq} (upper part), that an explosion energy is increasing smoothly for larger fields at $B_0 \gtrsim 6\cdot 10^{11}$ G. For smaller values of $B_0$ this dependence becomes chaotic, and is concentrated around a limiting curve, which seems to be independent on $B_0$. Such behaviour could be connected with a magnetic field amplification by action of a specific type of a magnetorotational instability \cite{tayler1973, spruit1999} (a Tayler-type instability, i.e. the one with a dominance of the toroidal magnetic field), appearing in earlier calculations \cite{mrsn2005, mrsn2006}, together with a magnetized matter fall-back after the core bounce.
This instability is developing at smaller $B_0$, and is characterised by rapid growth of the chaotic magnetic field, which finally exceeds $B_0$. To illustrate the field growth for low-to-intermediate field cases, the evolution of the poloidal magnetic energy in the PNS for the northern and southern hemispheres is shown for $DQ$ model with $B_0=5\cdot 10^{10}$ G in Figure \ref{fig:dq5e10_b_evol}.

At first $200-250$ ms after the core bounce the poloidal field smoothly grows in the regions within $r < 30$ km in both hemispheres due to the magnetized matter fallback from the exterior parts of the core. After $250$ ms, the field amplifies rapidly within $10$ km $ < r < 20$ km from the star's centre (see the red curves of Fig. \ref{fig:dq5e10_b_evol}). The toroidal component becomes dominant in the field structure soon after the bounce, and it is more likely, that the poloidal field is growing at $t_{p.b.} \gtrsim 250$ ms in both hemispheres due to the Tayler-type instability. This instability may be clearly seen in the exponential growth of the poloidal field in the southern hemisphere (lower part of Fig.\ref{fig:dq5e10_b_evol}). The resulting northern poloidal field remains stronger than the southern one in all low-to-intermediate field models ($B_0 < 6\cdot 10^{11}$ G).  As a result, the wrapped toroidal field provides a stronger magnetic pressure gradient in the north. It leads to the stronger northern jet and the negative PNS velocity (see Fig. \ref{fig:expl_dq}). The field amplification in the models with low-to-intermediate magnetic fields saturates and starts to produce the MR explosions efficiently at $t_{p.b.} \sim 500$ ms.

\color{black}

\begin{figure}[!htp]
		\centering
        \includegraphics[width=8.6cm,height=7.5cm]{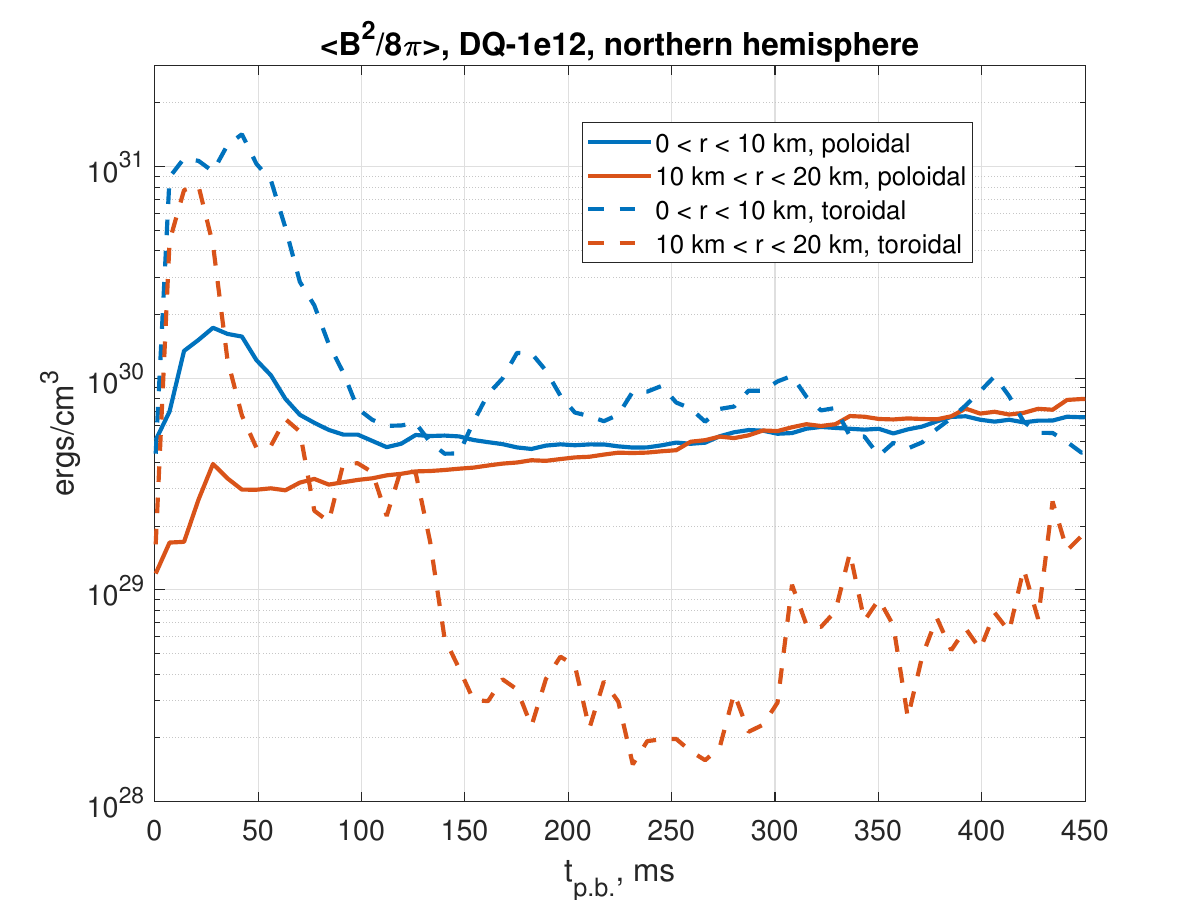}
		\includegraphics[width=8.6cm,height=7.5cm]{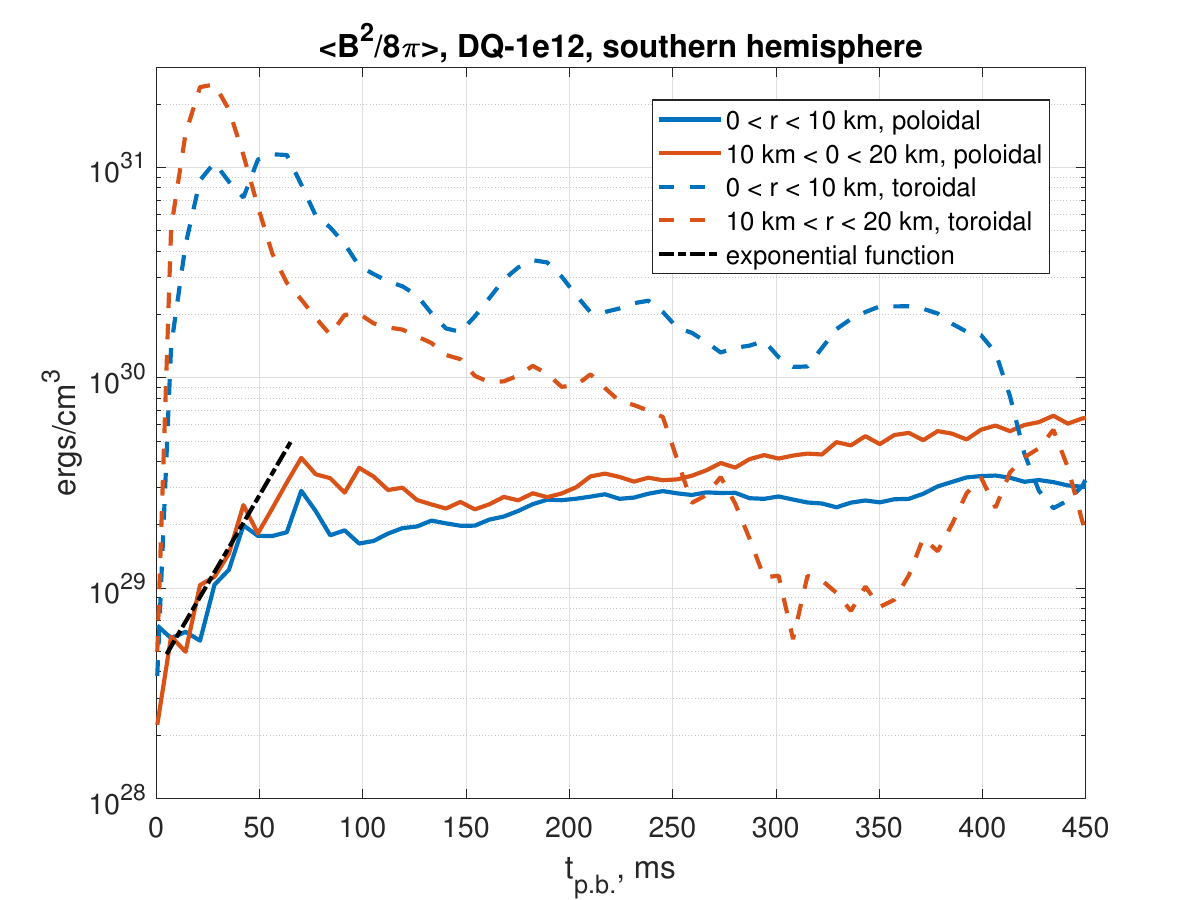}
		\caption{ Averaged density of poloidal (solid lines) and toroidal (dashed lines) magnetic energies in two spherical shellular regions within $0 < r < 10$ km (blue) and $10$ km $ < r < 20$ km (red) inside the PNS as the functions of the post-bounce time for the model $DQ-1e12$. Upper panel -- northern hemisphere, lower panel -- southern hemisphere. An exponential function is plotted (a black dash-dotted line) in the lower panel for the region with a rapid growth of the poloidal field. }
		\label{fig:dq1e12_b_evol}
	\end{figure}

\color{black}
The direction of the PNS kick is changed for the largest magnetic fields in the models $DQ-6e11$ and $DQ-1e12$. For the model with $B_0 = 10^{12}$ G the field amplification with time is plotted in Fig. \ref{fig:dq1e12_b_evol}. In this case the northern toroidal field induced by winding influences the angular velocity, decreasing it. The northern jet starts at $\sim 20$ ms p.b. due to the high magnetic pressure gradient in the region of the maximal angular velocity gradient at $\sim 10$ km from the star's centre. In the corresponding southern region, the poloidal field is growing rapidly during the first $70$ ms p.b., leading to a subsequent growth of the toroidal field by wrapping (see the lower panel of Fig. \ref{fig:dq1e12_b_evol}). The magnetic field growth inside $r \lesssim 20$ km leads to the southern jet launching at $\sim 80$ ms p.b., which becomes more intensive, than the northern one. According to Fig. \ref{fig:expl_dq} (black solid lines), at $t \gtrsim 200$ ms p.b. we observe the stronger southern outflow and the positive PNS kick velocity. 
The exponential growth of the magnetic field due to action of a dominant toroidal field in the differentially rotating region was studied by H.Spruit \cite{spruit1999}. The exponential function is plotted in the lower panel of Fig. \ref{fig:dq1e12_b_evol} in order to illustrate the rapid field growth. 

\color{black}

\begin{figure*}[!htp]
		\centering
		\includegraphics[width=5.0cm,height=7.5cm]{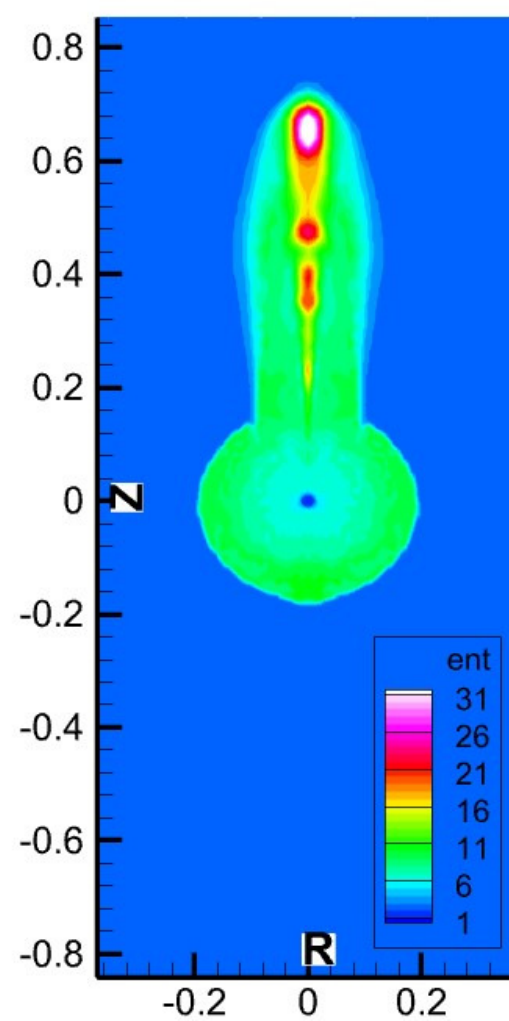}
		\includegraphics[width=5.0cm,height=7.5cm]{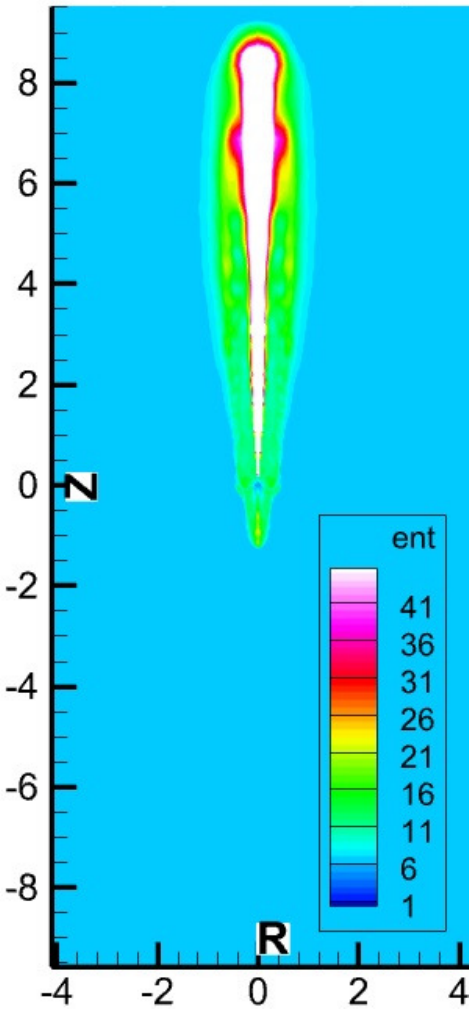}
		\includegraphics[width=5.0cm,height=7.5cm]{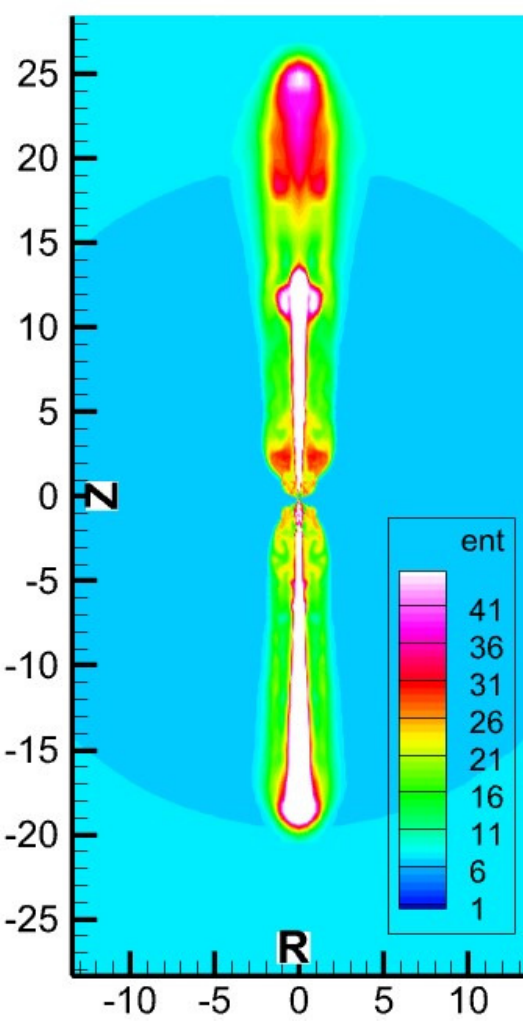}
		\caption{ Specific entropy (in units $k_B/m_u$) for the model $DQ-3e11$ at different times: left panel -- $t_{p.b.} = 84$ ms, central panel -- $t_{p.b.} = 350$ ms, right panel --  $t_{p.b.} = 896$ ms. A spatial axes scale is in units of 1000 km. }
		\label{fig:entropy_dq}
	\end{figure*}

In Fig.\ref{fig:entropy_dq} the time evolution of the entropy distribution is given for the initial magnetic field $B_0=3\cdot 10^{11}$ G. For this case the MR mechanism start to launch the jets at approximately $\sim 80$ ms after the core bounce. The resulting northern jet is stronger. On the right panel the entropy of the southern jet is larger because of its lower density. All models show jet-like structures, with high entropy and a moderate collimation for low magnetic field. The jet collimation is growing with increasing of $B_0$. For $B=10^{12}$ G we observe a more energetic southern jet, unlike to the cases with lower magnetic fields.

\begin{figure*}[!htp]
		\centering  
        \includegraphics[width=5.8cm,height=4.9cm]{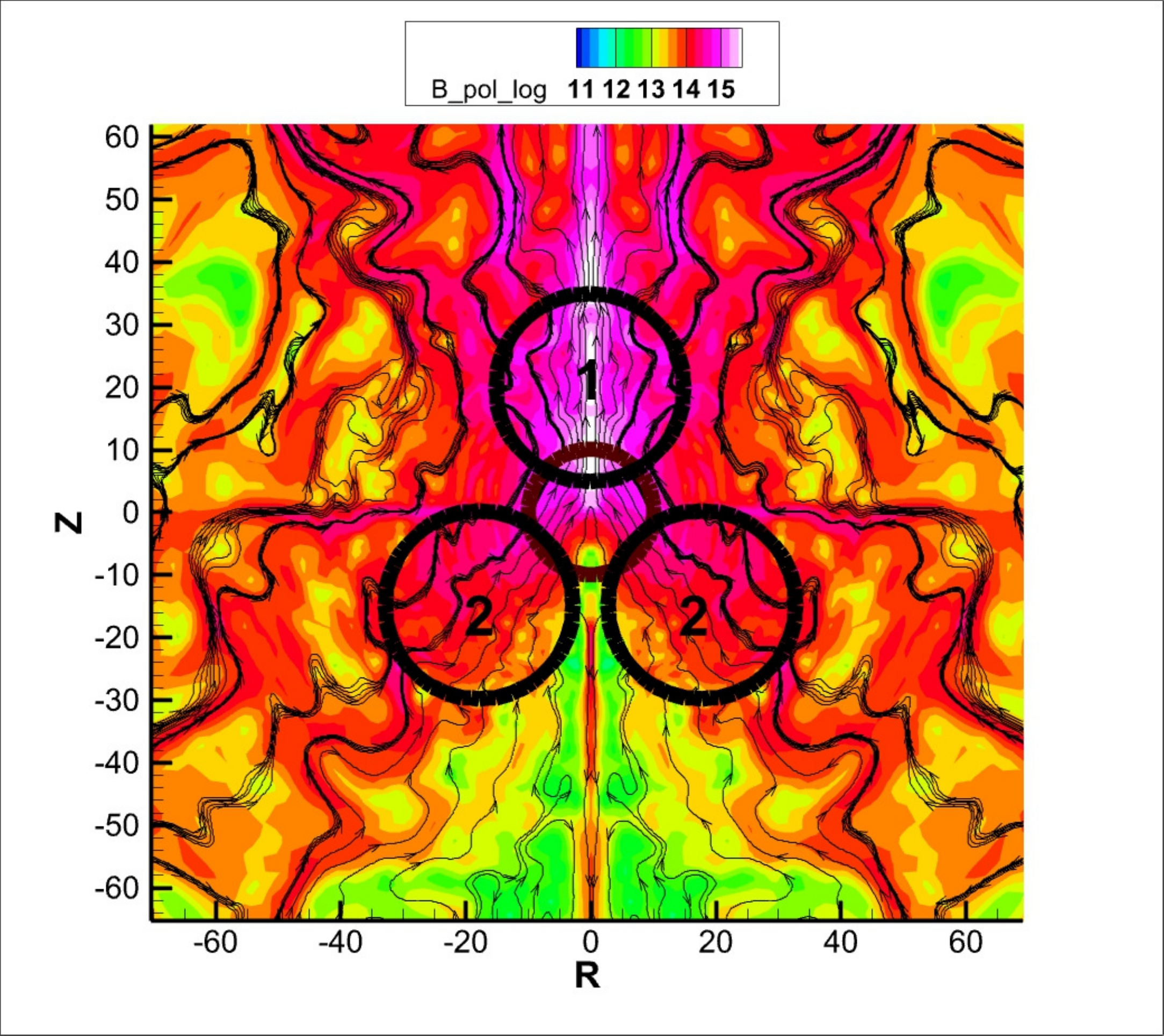}
		\includegraphics[width=5.8cm,height=4.9cm]{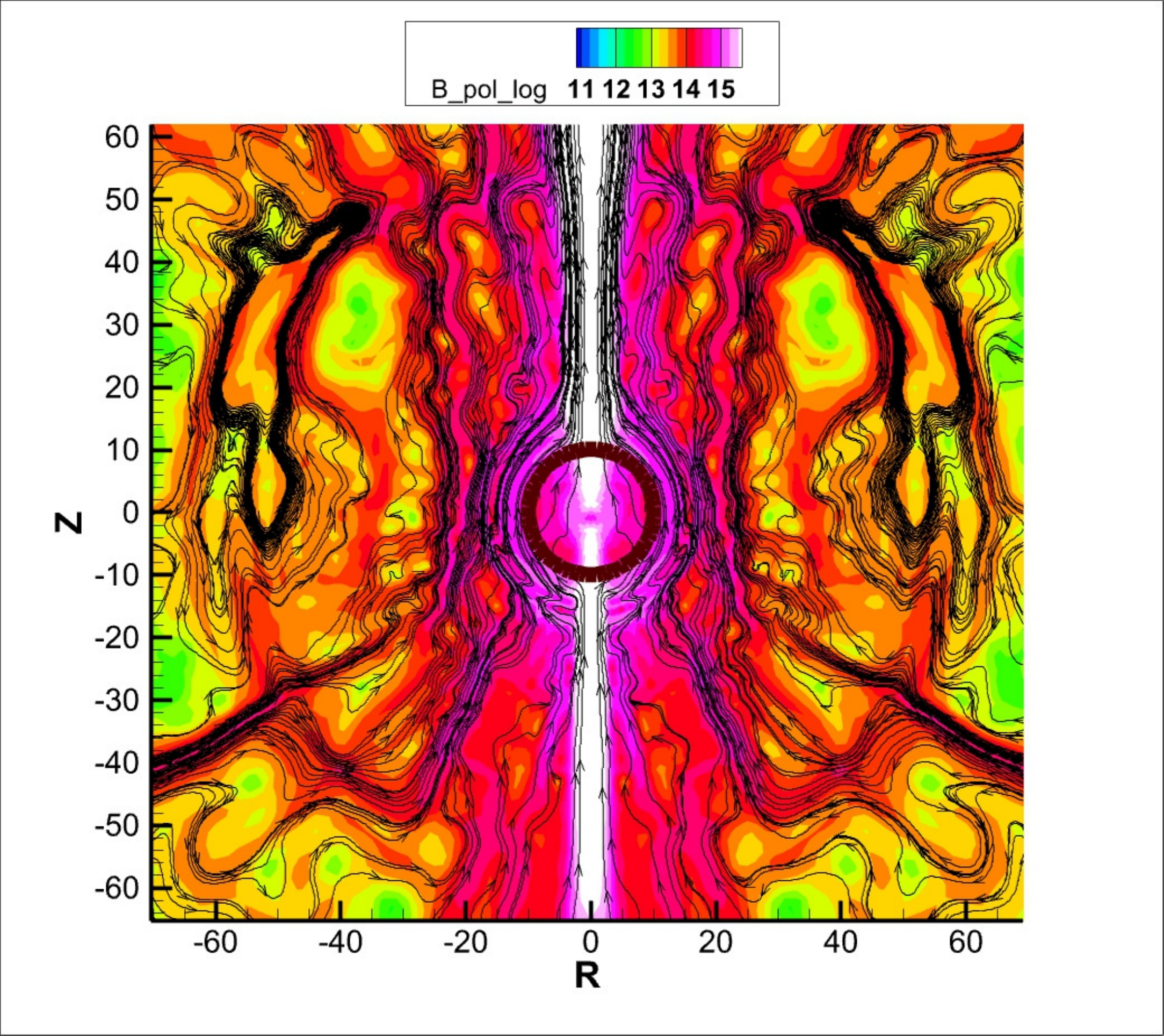}
		\includegraphics[width=5.8cm,height=4.9cm]{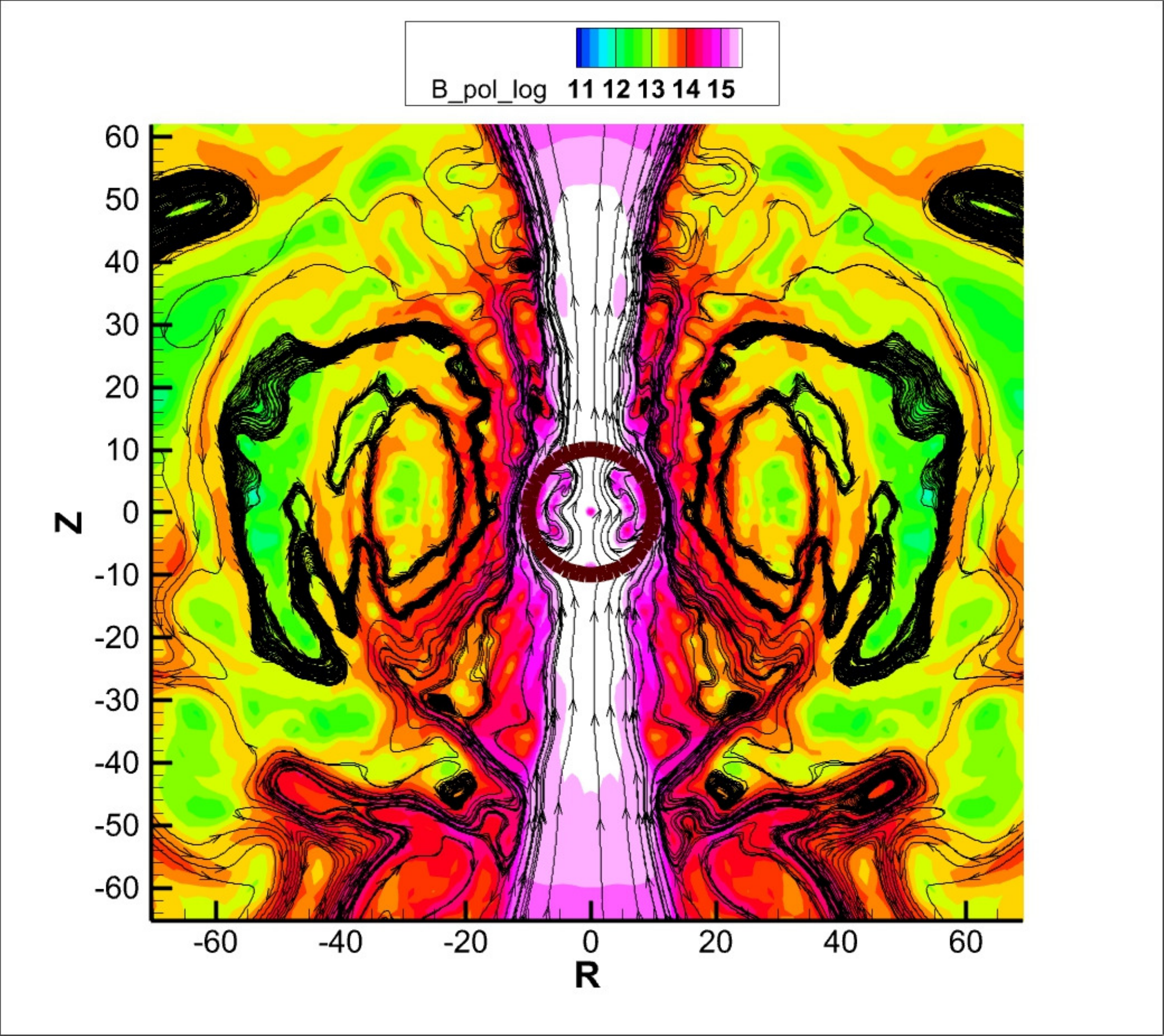}
		\caption{ A logarithm of poloidal magnetic field absolute value (Gauss) in the PNS region for $DQ$ model with $B_0=3\cdot 10^{11}$ G, at different post-bounce times: left panel:  $t_{p.b.} = 14$ ms, central panel:  $t_{p.b.} = 350$ ms, right panel:  $t_{p.b.} = 896$ ms. Black curves  show  poloidal field lines. Circle in the center  marks a radius of 10 km. The distance from the centre of the star in cylindrical coordinates is plotted on the axes in units of kilometres. On the left panel  the dipolar and quadrupolar parts of the magnetic field are summed up near the north pole region $1$. In regions $2$ the field lines are more complicated due to more pronounced start of chaotic motion. }
		\label{fig:bfield_dq}
	\end{figure*}

The logarithm of the poloidal field absolute value in the vicinity of the PNS region in the model $DQ-3e11$ is illustrated in Fig. \ref{fig:bfield_dq} for different post-bounce times. After the bounce, we see an asymmetric structure of the poloidal field (Fig. \ref{fig:bfield_dq}, left panel), with a stronger field above the equator, as the precollapse field. At later stages, the chaotic motions of the fluid inside the region with radius $\sim 50$ km, connected with a possible development of the magnetorotational instability together with a convective motion distort the poloidal field. In regions $2$ the field lines are more complicated due to more pronounced start of chaotic motion.
\color{black}
The field remains stronger in the upper hemisphere, producing a stronger jet in a northern direction with a southern velocity kick. 
\color{black}

\color{black}
In conclusion to $DQ$ models analysis, the following dynamics of the simulated MR explosions is established. For low-to-intermediate initial magnetic fields, the field amplification is developing in both hemispheres, and the values of resulting fields follow the values of the initial fields. Therefore, the gradient of the magnetic pressure in the north remains stronger, and the forming jet is also stronger in the northern direction. For high values of $B_0$ the poloidal field is not amplified above the equator, while a rapid field growth is pronounced below the equatorial plane, resulting in the stronger outflow there. Such changing of the jet asymmetry and accompanying kick velocity of PNS on the value of the initial magnetic field are clearly visible in the Table \ref{tab:table_results_dq}.  
\color{black}

\subsection{Results for the offset dipole fields}

The case  of the offset dipole in the core before collapse is qualitatively close to the one with a composition of dipolar and quadrupolar fields, and three simulated models in this subsection show a similar dynamics. This fact is expected, because the offset configuration can be expanded in a multipole series, with a larger contribution of higher multipoles with increasing of the offset parameter $z_{off}$. As in $DQ-$ models, we obtain jet-like explosions with different intensities in the southern and the northern hemispheres. 

\color{black} 
We have obtained for $Do-$ models, that the direction of the kick is changed with changing of the value of the initial field $B_0$. Therefore, the resulting jet is stronger from the hemisphere with stronger initial magnetic energy for lower values of $B_0$, and changes its direction for the largest $B_0$, like in $DQ$ models. \color{black}
 
	\begin{table}[h!]
		\begin{center}
			\caption{The simulation results for the models with a north offset dipolar magnetic field. Explosion energies, PNS kick velocities and anisotropy parameter $A_E$ are shown. The final post-bounce times are given in the last column. }
			\label{tab:table_results_od}
			\begin{tabular}{c|c|c|c|c}
				\textbf{model} & $E_{expl}$, $10^{51}$ ergs & $v_{kick}$, km/s & $A_E$ & $t_{p.b.}^f$, ms \\
				\hline
				$Do-3e10$ & 0.599 & -337.9 & 0.683 & 1300 \\
				$Do-2e11$ & 0.457 & -124.5 & 0.332 & 1188 \\
				$Do-1e12$ & 1.007 &  115.5 & -0.151 & 1073 \\
			\end{tabular}
		\end{center}
	\end{table}
	
\begin{figure}[!htp]
		\centering
        \includegraphics[width=8.6cm,height=6.0cm]{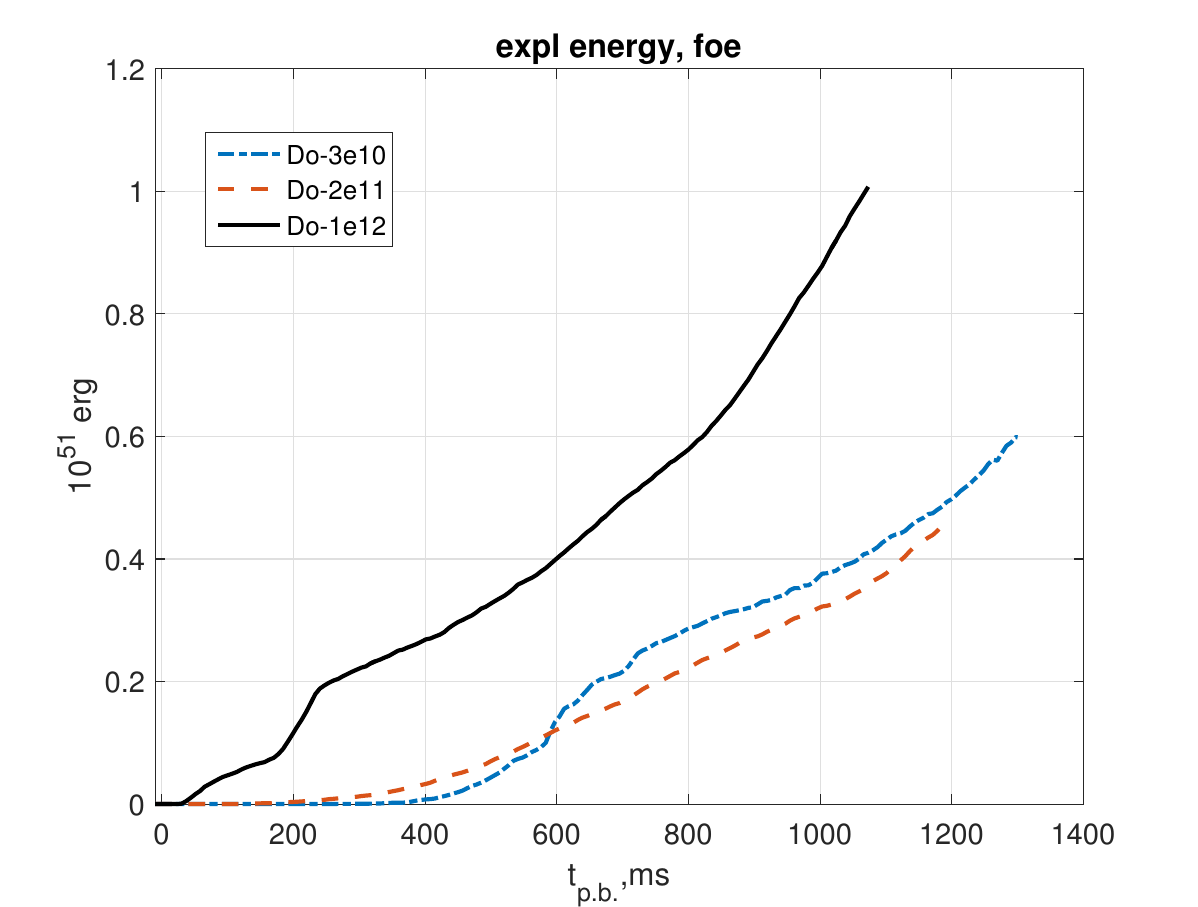}
		\includegraphics[width=8.6cm,height=6.0cm]{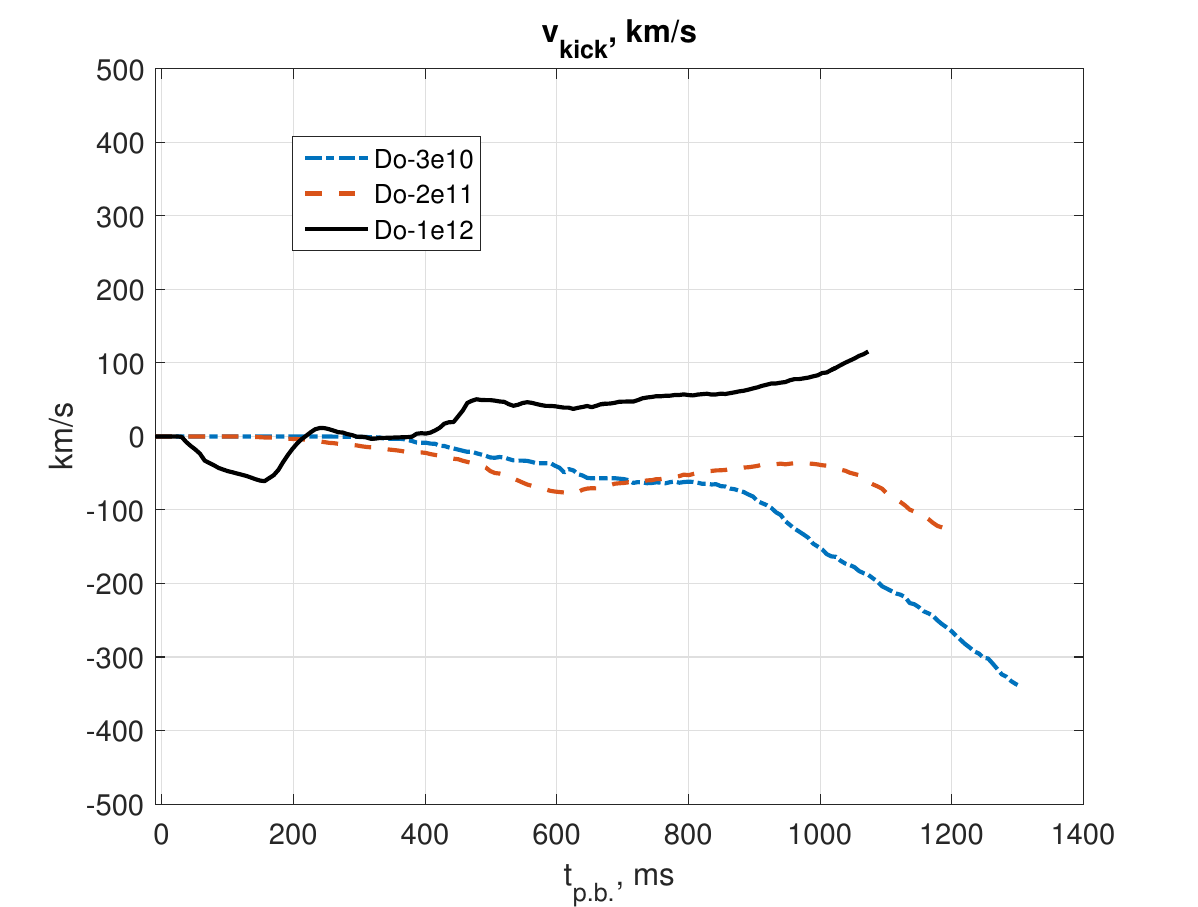}
		\caption{ Integral characteristics for configurations with offset dipolar magnetic fields as functions of the post-bounce time: upper panel gives explosion energy, lower panel gives PNS kick velocity. }
		\label{fig:expl_od}
	\end{figure}

In all models, the jet appears earlier in the part, where the magnetic field value was larger at the onset of the collapse (northern side). The energies, the PNS kicks, and the explosion anisotropy values, are presented for $Do$ models in the Table \ref{tab:table_results_od}. Figure \ref{fig:expl_od} shows the energies and PNS kicks dependence on the post-bounce time. 
	
The  parameter $r_0$ in \eqref{magfield}, characterizing a field localization, by computational reasons, was taken  smaller in the $Do$ models, than in the corresponding $DQ$ ones. As a result we start with a smaller amount of magnetic energy before the collapse, and explosions are weaker in $Do$ models at the same $B_0$. From comparison of Tables \ref{tab:table_results_dq} and \ref{tab:table_results_od} we see, that $Do$ kick velocities are lower at smaller $B_0$, than for $DQ$. At smaller $B_0$ the explosion asymmetry produces more energetic northern jet. At largest calculated $B_0 = 10^{12}$ G the southern jet becomes stronger, like in $DQ$ models, compare Figs.\ref{fig:expl_dq} and \ref{fig:expl_od}.  

	\begin{figure*}[!htp]
		\centering
		\includegraphics[width=5.0cm,height=7.5cm]{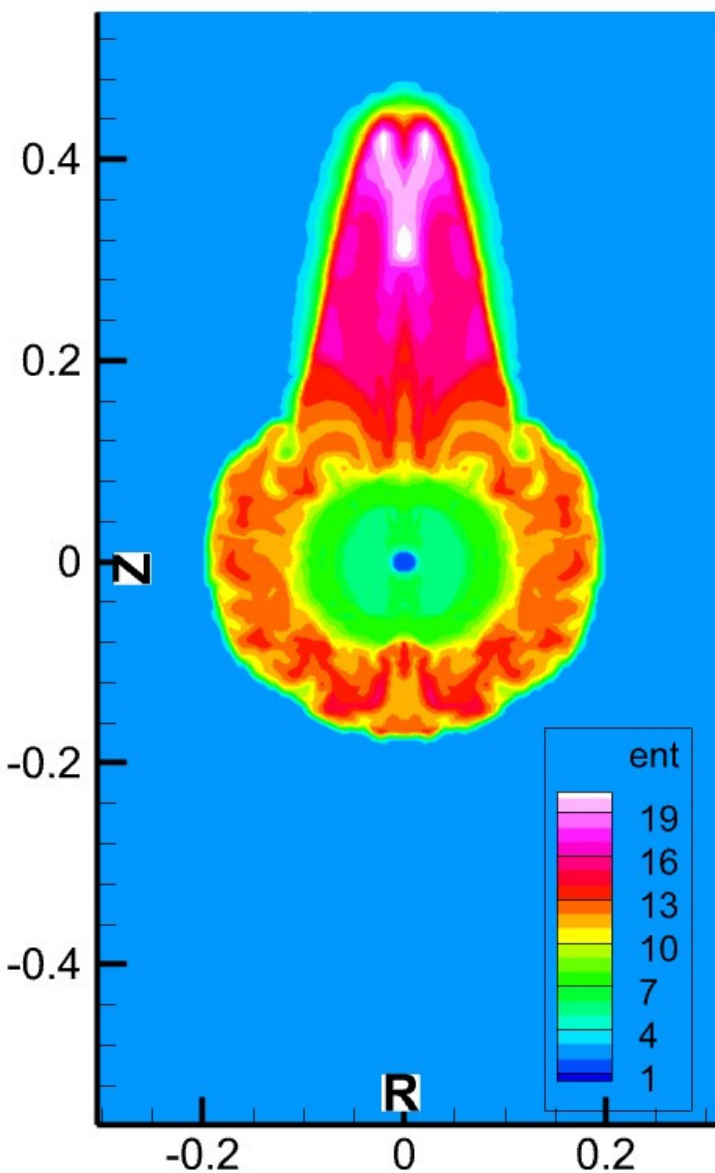}
		\includegraphics[width=5.0cm,height=7.5cm]{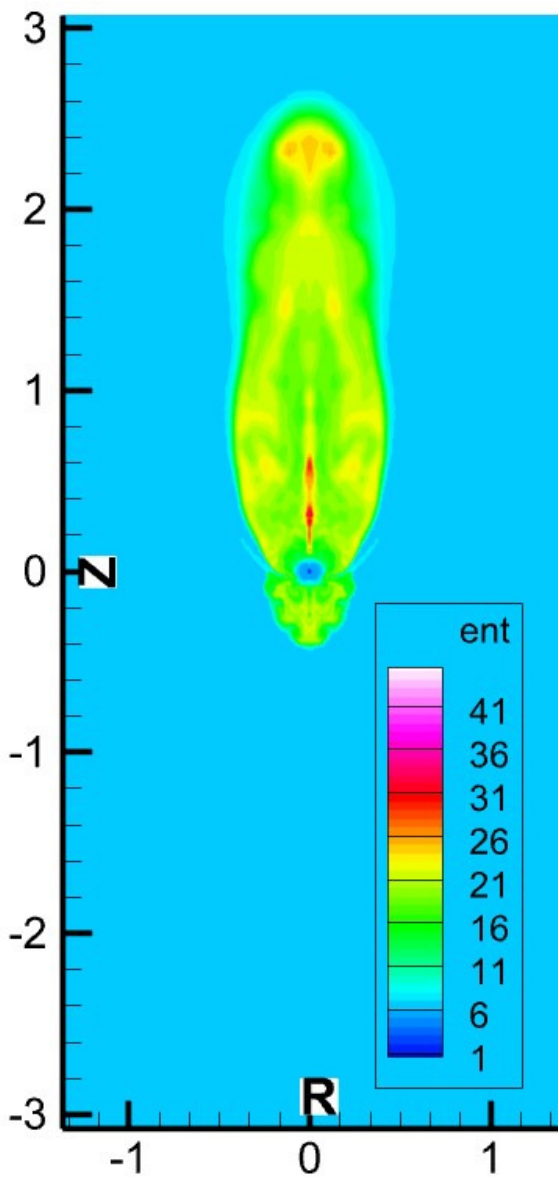}
		\includegraphics[width=5.0cm,height=7.5cm]{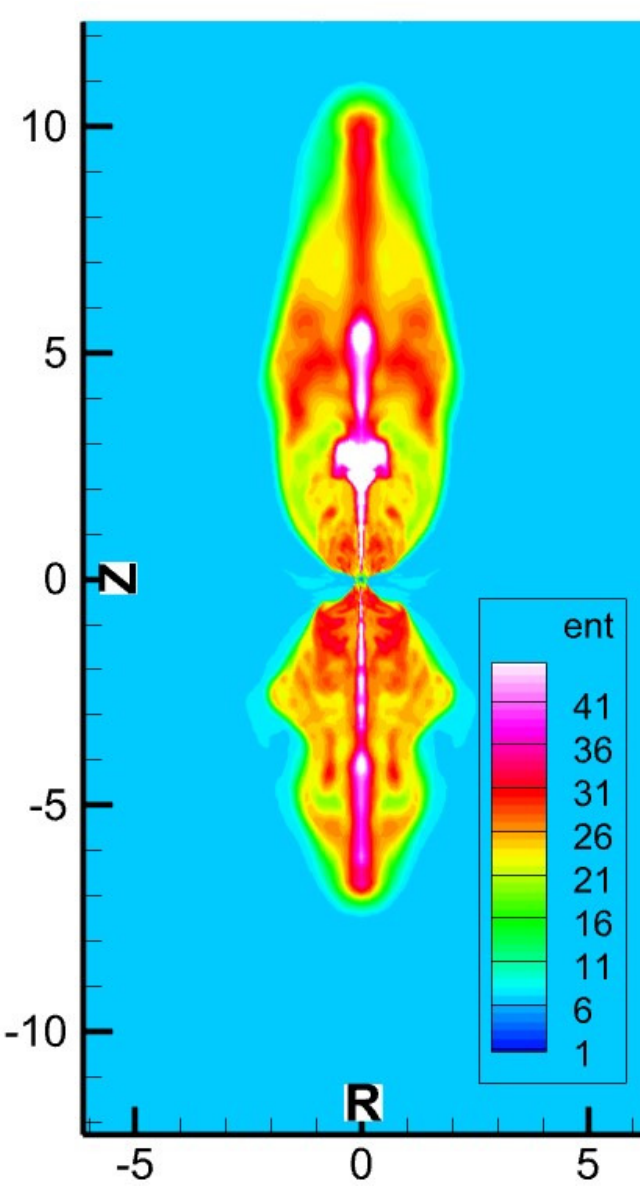}
		\caption{ Specific entropy (in units $k_B/m_u$) for the model $Do-3e10$ at different times: left panel -- $t_{p.b.} = 233$ ms, central panel -- $t_{p.b.} = 401$ ms, right panel --  $t_{p.b.} = 800$ ms. A spatial axes scale is in units of 1000 km. }
		\label{fig:entropy_od}
	\end{figure*}

Appearance of the kick in strongly magnetized PNS had been investigated in \cite{sawai2008} for initial dipole field $B_{0}$ around $5\cdot 10^{13}$ G. Formation of the kick was connected with an offset position of the dipole. In all variants a stronger jet had a direction along the offset shifting, from the half part of PNS having larger magnetic energy. This result is different from  our calculations, see Fig.\ref{fig:expl_od} and Tab.\ref{tab:table_results_od}. The difference could be connected with very high initial magnetic field in the paper \cite{sawai2008}. It had suppressed all instabilities, developing in our calculations, and changing the direction of the kick. In our calculations with weaker fields the instabilities are developing in both hemispheres, resulting in a stronger jet in the northern direction from the largest field hemisphere (see the entropy distribution of the outflows for the model $Do-3e10$ in Fig. \ref{fig:entropy_od}). In our case of strong field $Do-1e12$ the instability is not developing in the northern hemisphere with a larger field, but is developing in the opposite hemisphere, amplifying the field there and producing the stronger jet in the opposite direction. \color{black} The magnetic field evolution for the strongest field case below and above the equator during the first $100$ ms after the bounce provides the same features, as in the corresponding model with dipole+quadrupole fields, see Fig. \ref{fig:dq1e12_b_evol} for $DQ-1e12$. \color{black}

\subsection{Results for the superposition of dipolar and toroidal fields}

The case of initial superposition of dipolar field and a strong toroidal component proceeds differently from the first two options. The asymmetry is not prescribed here from the beginning, but appears in a differentially rotating star because of initial difference in a symmetry types of poloidal and toroidal components \cite{BKM1992}. Violation of the mirror symmetry after beginning of the collapse, in presence of a differential rotation, is  consisted of increasing of the toroidal field more rapidly in  one (southern) hemisphere, and slower in another (northern) one.

In all $DT$- type models the initial strength of toroidal field was taken as $B_{0,tor} = 10^{13}$ G in \eqref{magfield}.
 
    \begin{table}[h!]
        \begin{center}
        \label{tab5t}
		  \caption{Simulation results for models with a superposition of dipole and toroidal fields. Explosion energies, PNS kick velocities and anisotropy values $A_E$ are shown. The last post-bounce time moments are given in the right column. }
		      \label{tab:table_results_dt}
			\begin{tabular}{c|c|c|c|c}
				\textbf{model} & $E_{expl}$, $10^{51}$ ergs & $v_{kick}$, km/s & $A_E$ & $t_{p.b.}^f$, ms \\
				\hline
			  
				$DT-4e10$ & 0.559 & -256.5 & 0.528 & 1300 \\
				$DT-6e10$ & 0.532 & -279.1 & 0.577 & 1200 \\
				$DT-1e11$ & 0.521 & -212.4 & 0.448 & 1052 \\
				$DT-3e11$ & 0.682 & -56.21 & 0.040 & 1038 \\
				$DT-6e11$ & 1.351 & -14.10 & -0.014 & 1024 \\
				$DT-1e12$ & 2.269 &  10.57 & -0.035 & 1015 \\
				
			\end{tabular}
	    \end{center}
    \end{table}

\begin{figure}[!htp]
		\centering
        \includegraphics[width=8.6cm,height=6.0cm]{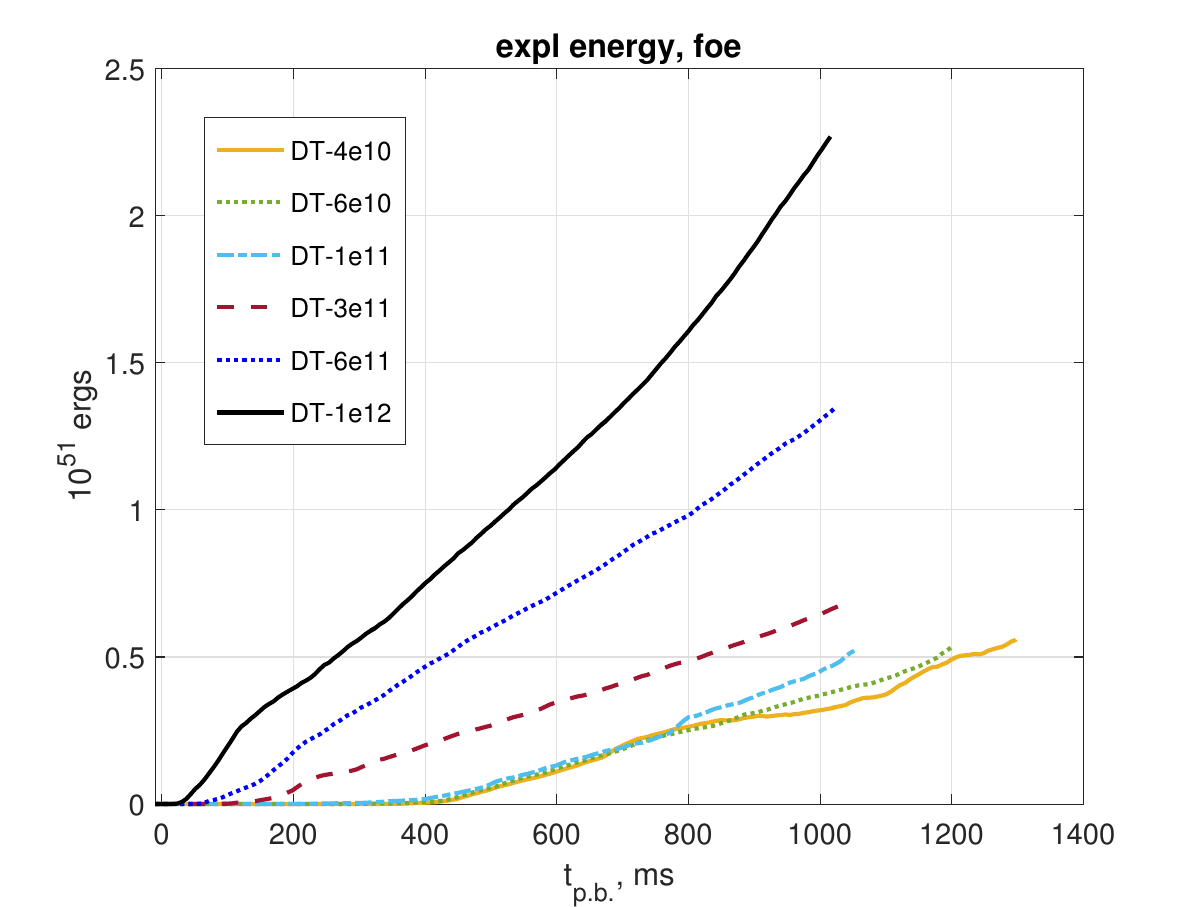}
		\includegraphics[width=8.6cm,height=6.0cm]{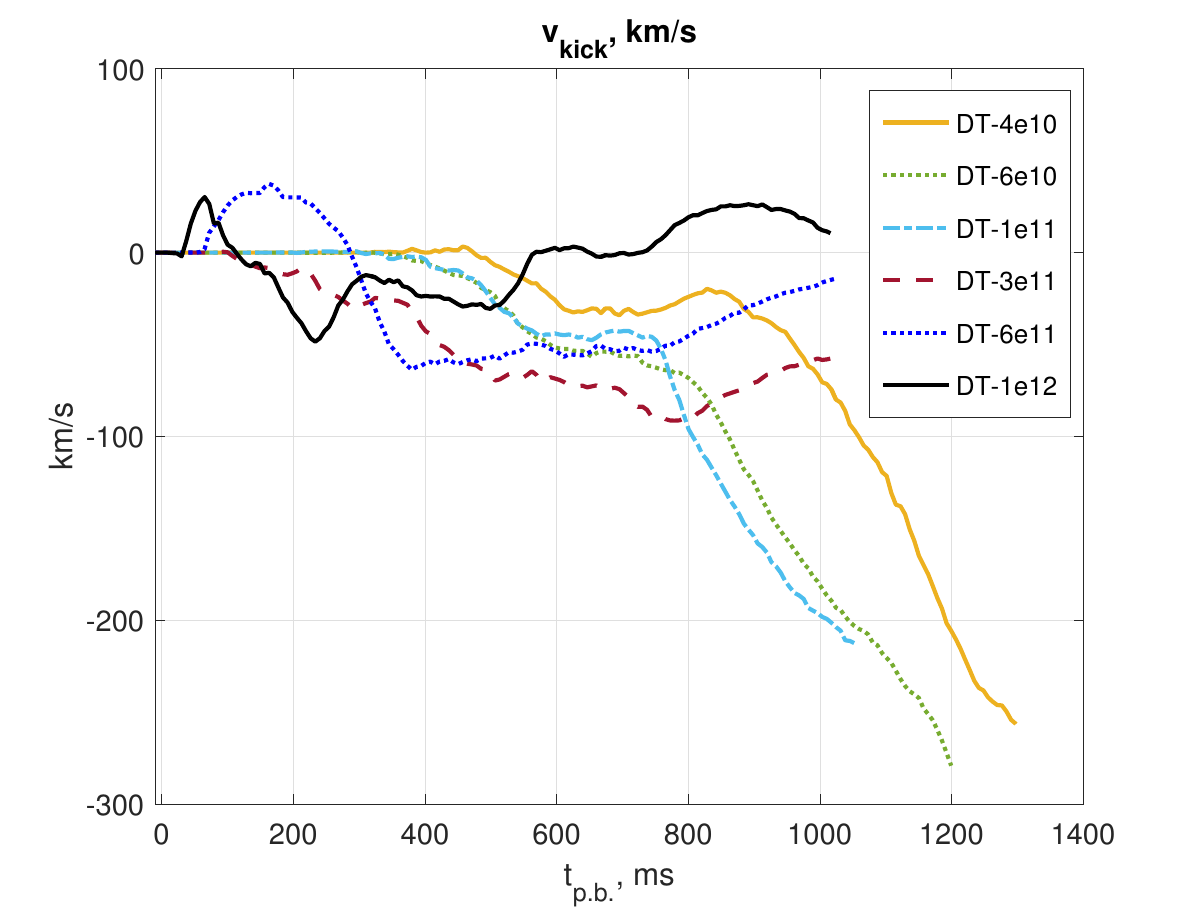}
		\caption{ Integral characteristics for configurations with the superposition of dipolar and toroidal fields as the functions of the post-bounce time: upper panel -- explosion energy, lower panel -- PNS kick velocity. }
		\label{fig:expl_dt}
	\end{figure}

The energies, PNS kicks and explosion anisotropy values are presented in Table \ref{tab:table_results_dt}. Figure \ref{fig:expl_dt} shows post-bounce time evolution of the energies and PNS kicks velocities. One can distinguish the models into two groups. The first group of the models with intermediate initial dipole fields $4e10,\,\,6e10,\,\,1e11$ G in Table \ref{tab:table_results_dt},  provides similar explosion properties with intermediate values of asymmetry and south direction kick velocities. In the second group of models with high magnetic fields $3e11,\,\,6e11,\,\,1e12$ G in Table \ref{tab:table_results_dt}, we have obtained a decrease of asymmetry of the blast waves and disappearance of the PNS kick. In the last models $DT-6e11$ and $DT-1e12$ the PNS kick velocity is inside the error box of numerical precision, and they may be considered as symmetric explosions without kick.

\begin{figure*}[!htp]
		\centering

        \includegraphics[width=5.8cm,height=4.9cm]{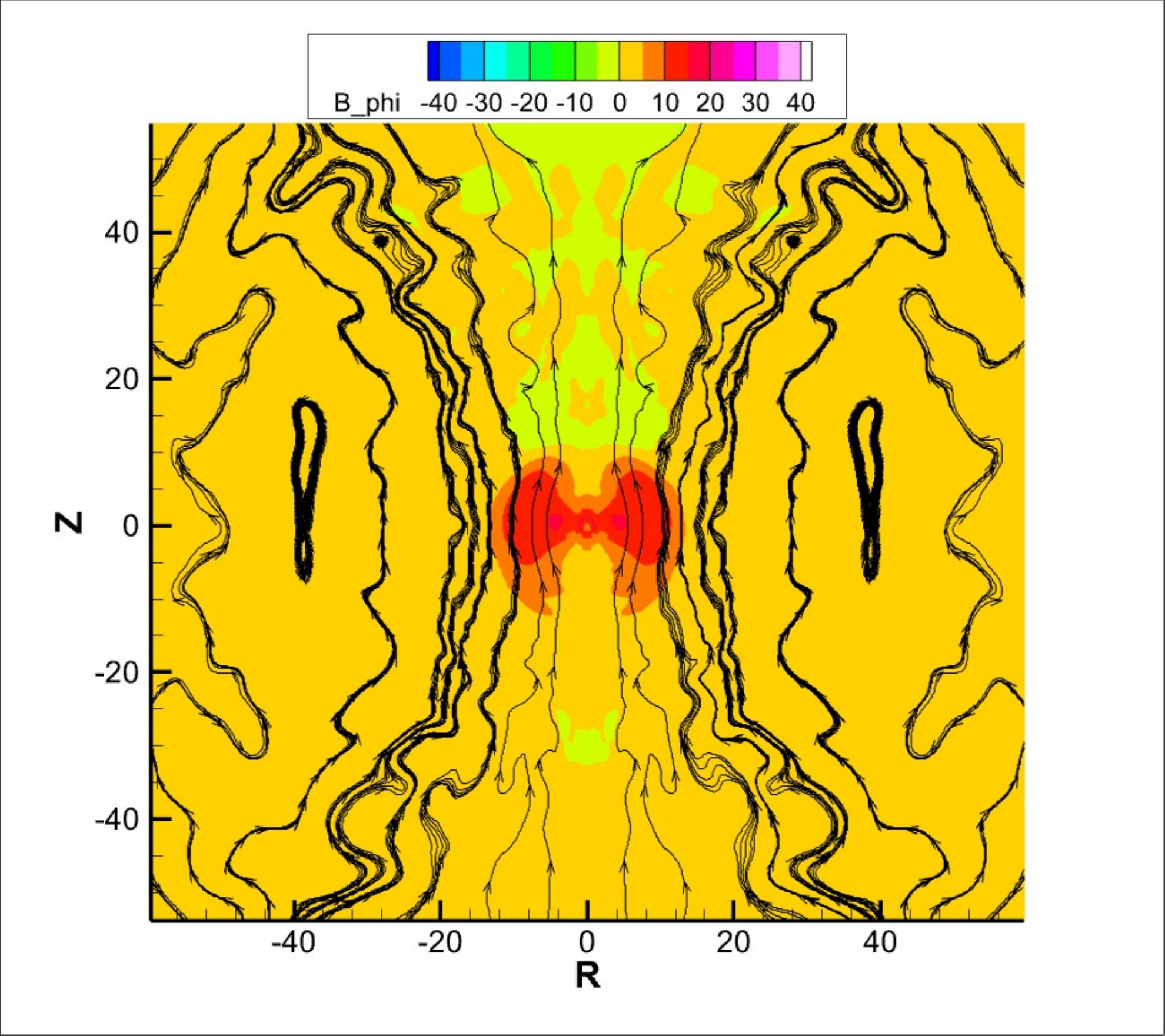}
		\includegraphics[width=5.8cm,height=4.9cm]{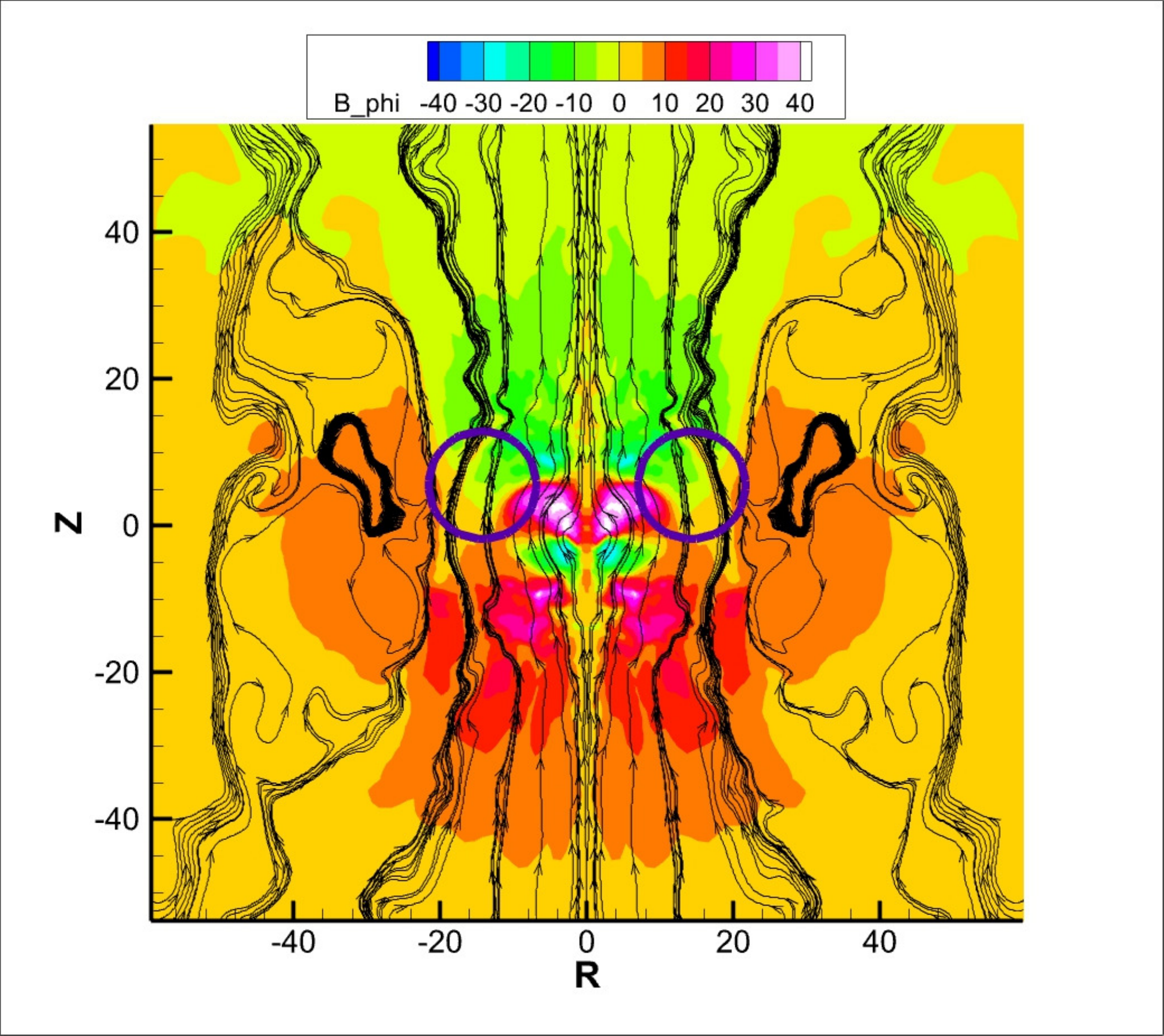}
		\includegraphics[width=5.8cm,height=4.9cm]{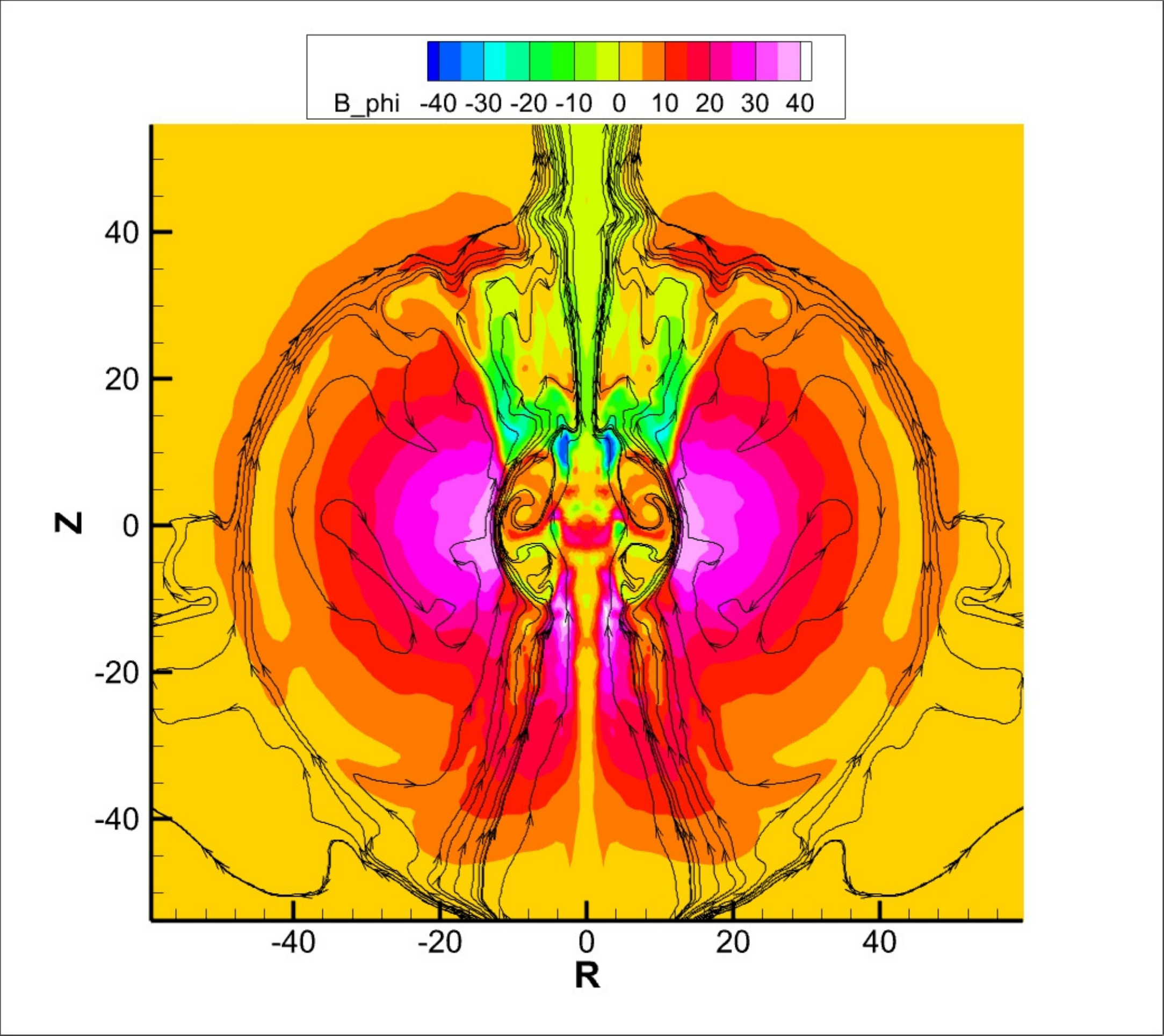}
  
		\caption{ A magnetic field distribution in the vicinity of the PNS at different post-bounce times in the model $DT-6e10$ (Left panel -- 9 ms p.b., central panel -- 254 ms p.b., right panel -- 646 ms p.b.). The distance from the centre of the star in cylindrical coordinates is plotted on the axes in kilometres, while the colour palette corresponds to the toroidal field value in the units of $10^{15}$ G. Black contours correspond to the lines of the poloidal magnetic field. The region with a highest gradient of the magnetic pressure is located in the northern hemisphere, it is marked with the circles at the central panel. }
		\label{fig:b_phi_dt}
	\end{figure*}

The models from the first group show similar properties of the explosion dynamics. The evolution of the magnetic field is illustrated in Fig.\ref{fig:b_phi_dt} for the model $DT-6e10$. At the moment of core bounce, the toroidal field represents a compressed configuration inherited mainly from the progenitor toroidal field, where it has an even structure with respect to the equator (left panel of Fig. \ref{fig:b_phi_dt}). Near the bounce the poloidal field  acquires different forms in two regions. Inside the inner 10 km from the center the magnetic field is concave with respect to the star center, while outside it has a convex structure  (see also \cite{obergaulinger2014, matsumoto2020}). Wrapping of the anti-symmetric dipole field due to differential rotation produces antisymmetric toroidal component, which is increasing with time.  The antisymmetric generated toroidal field in the inner part ($r \le 10$ km) of the PNS and the one in the regions with $r > 10$ km  has different signs (central panel of Fig. \ref{fig:b_phi_dt}, $t_{p.b} = 254$ ms). The overall structure of the magnetic field resembles a $"$twisted-torus$"$ \cite{cr2013} configuration with a dominance of the toroidal component. The largest gradient of the magnetic pressure lies in the regions above the equator (see black circles in Fig. \ref{fig:b_phi_dt}), where $B_{\phi,compr}^{sym}$ and $B_{\phi,wind}^{asym}$ have the same sign inside the inner 10 km, and where  the toroidal field distribution passes through zero. It results in a sharper gradient of the toroidal magnetic pressure, in comparison to the similar region below the equator. The magnetic pressure gradient is stronger in the northern hemisphere, and hence, the northern jet-like outflow becomes stronger, than the southern one. In the models $DT-4e10$, $DT-6e10$ and $DT-1e11$ we have got the asymmetry values $A_E \sim 0.5$ and kick velocities $v_{kick} \sim 200-300$ km/s at the final times of our simulations.

In the models with the highest poloidal precollapse fields $DT-3e11$, $DT-6e11$ and $DT-1e12$, the generated antisymmetric toroidal field becomes stronger than the symmetric compressed toroidal field inside the PNS. The initially appeared  asymmetry of the field is too small for producing a noticeable jet asymmetry, 
which is smeared out by the action of the wrapped antisymmetric field, with almost equal modules of the field strength. 
As a result, almost symmetric explosions are formed, see Table \ref{tab:table_results_dt}. Small jet asymmetry and small kick velocity is expected also in the case, when the initial toroidal field is large enough, and the wrapped field violation of the symmetry is small.

	\begin{figure*}[!htp]
		\centering
		\includegraphics[width=5.0cm,height=7.5cm]{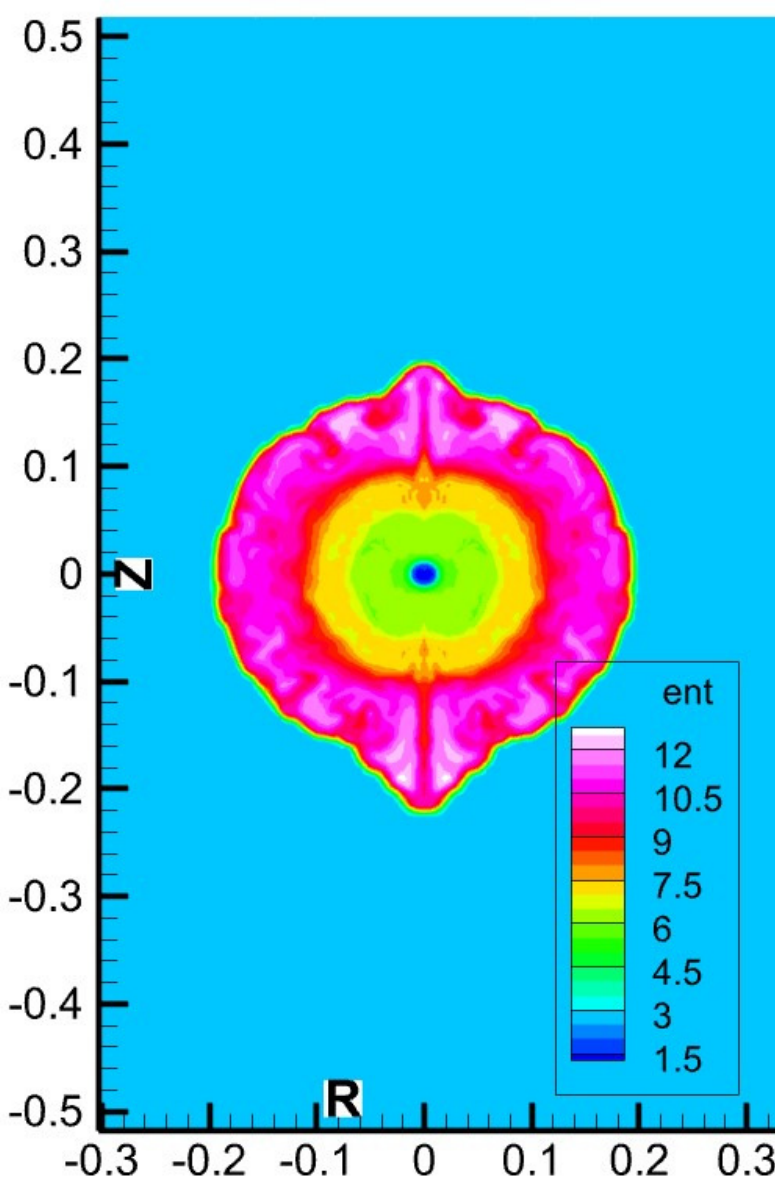}
		\includegraphics[width=5.0cm,height=7.5cm]{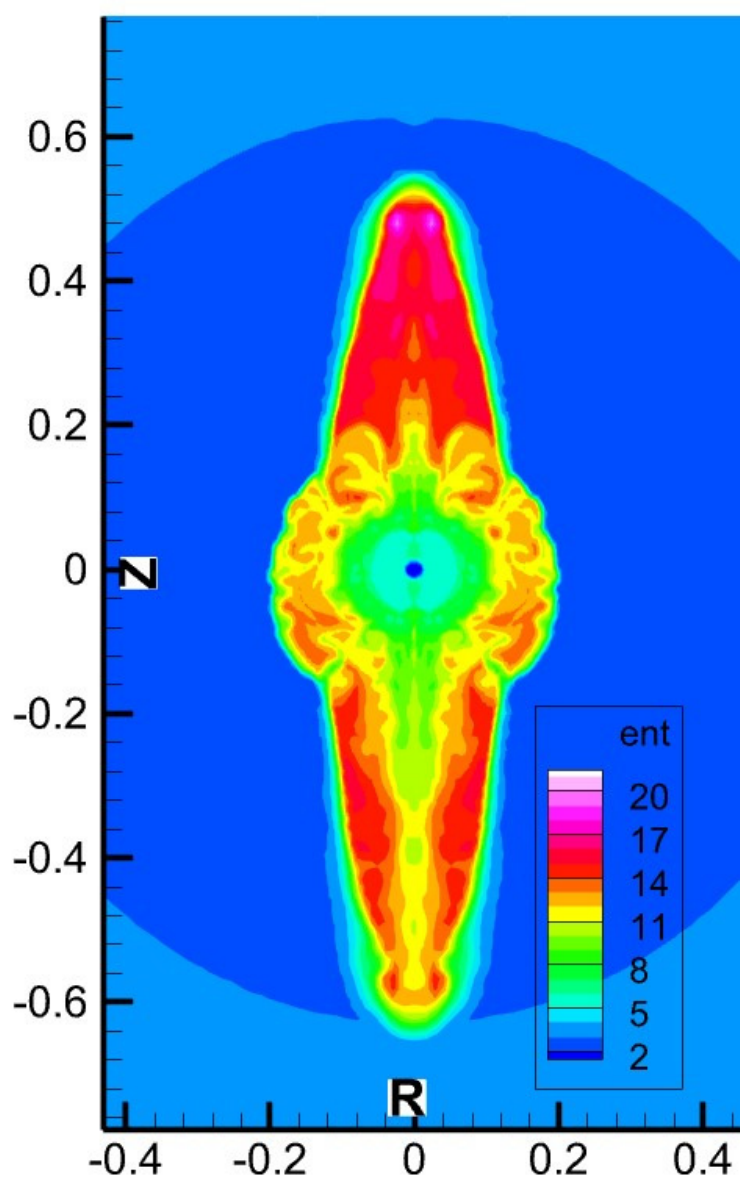}
		\includegraphics[width=5.0cm,height=7.5cm]{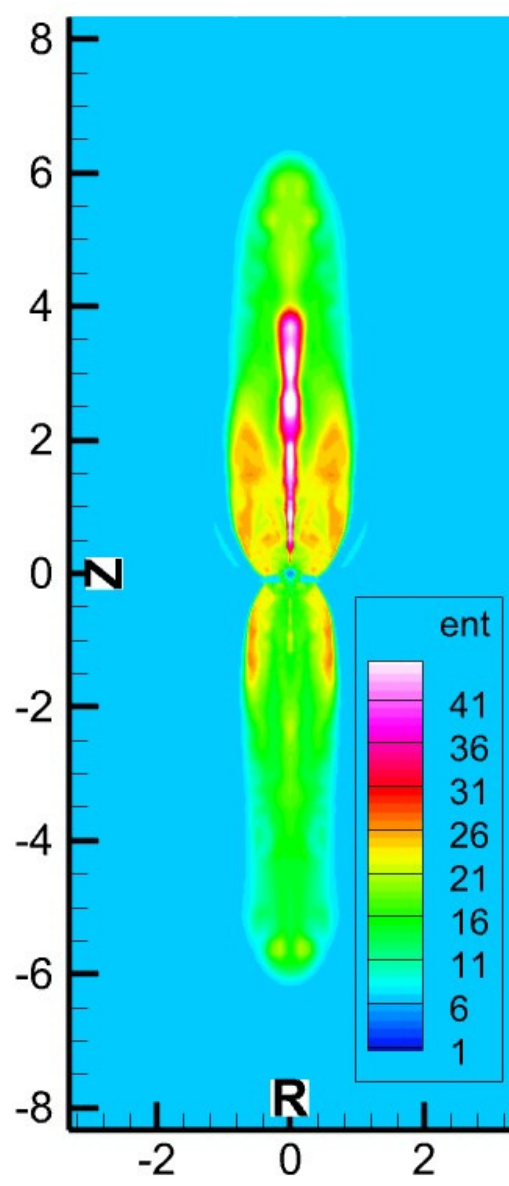}
		\caption{ Specific entropy (in units $k_B/m_u$) for the model $DT-6e10$ at different times: left panel -- $t_{p.b.} = 142$ ms, central panel --  $t_{p.b.} = 254$ ms, right panel --  $t_{p.b.} = 648$ ms. A spatial axes scale is in units of 1000 km. }
		\label{fig:entropy_dt}
	\end{figure*}

To illustrate the morphology of MR explosions in $DT-$ models, the specific entropy of the flow for the model $DT-6e10$ is plotted in Fig. \ref{fig:entropy_dt} at different stages of the explosion. 
At the beginning of the explosion stage, the jets are forming almost simultaneously and with only minor asymmetries (see left and central panels in Fig. \ref{fig:entropy_dt}). 
The  equatorial anisotropy of the explosion accumulates after they propagate over several thousands of kilometers from the centre (right panel of Fig. \ref{fig:entropy_dt}). \color{black} In this case the central engine starts to work anisotropically after beginning of the explosion phase, resulting in a development of the equatorial asymmetry of the outflows.
\color{black} Note, that for high precollapse magnetic fields from the second group, the shape of the ejected material remains almost symmetric with respect to the equator during the simulations. 

\begin{figure}[!htp]
		\centering
		\includegraphics[width=5cm,height=4cm]{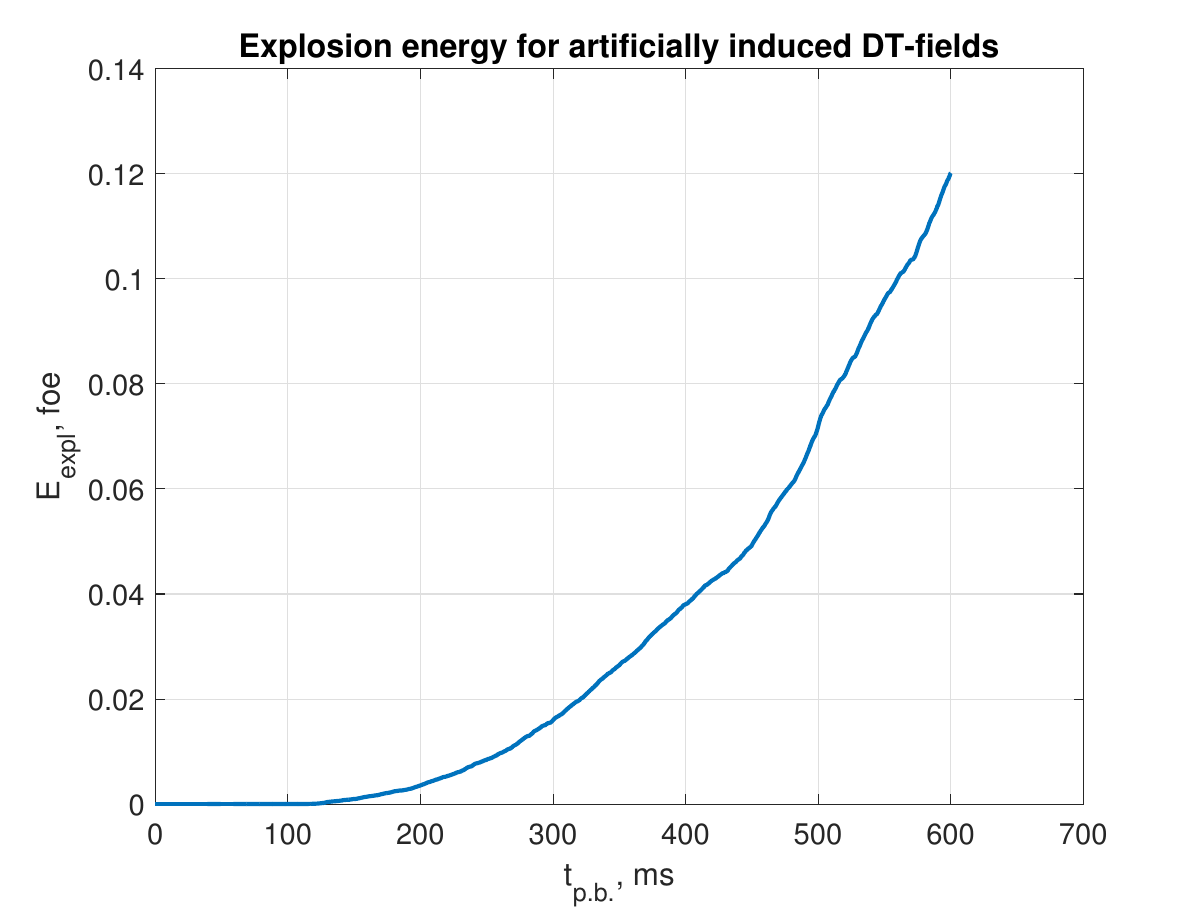}
		\includegraphics[width=5cm,height=4cm]{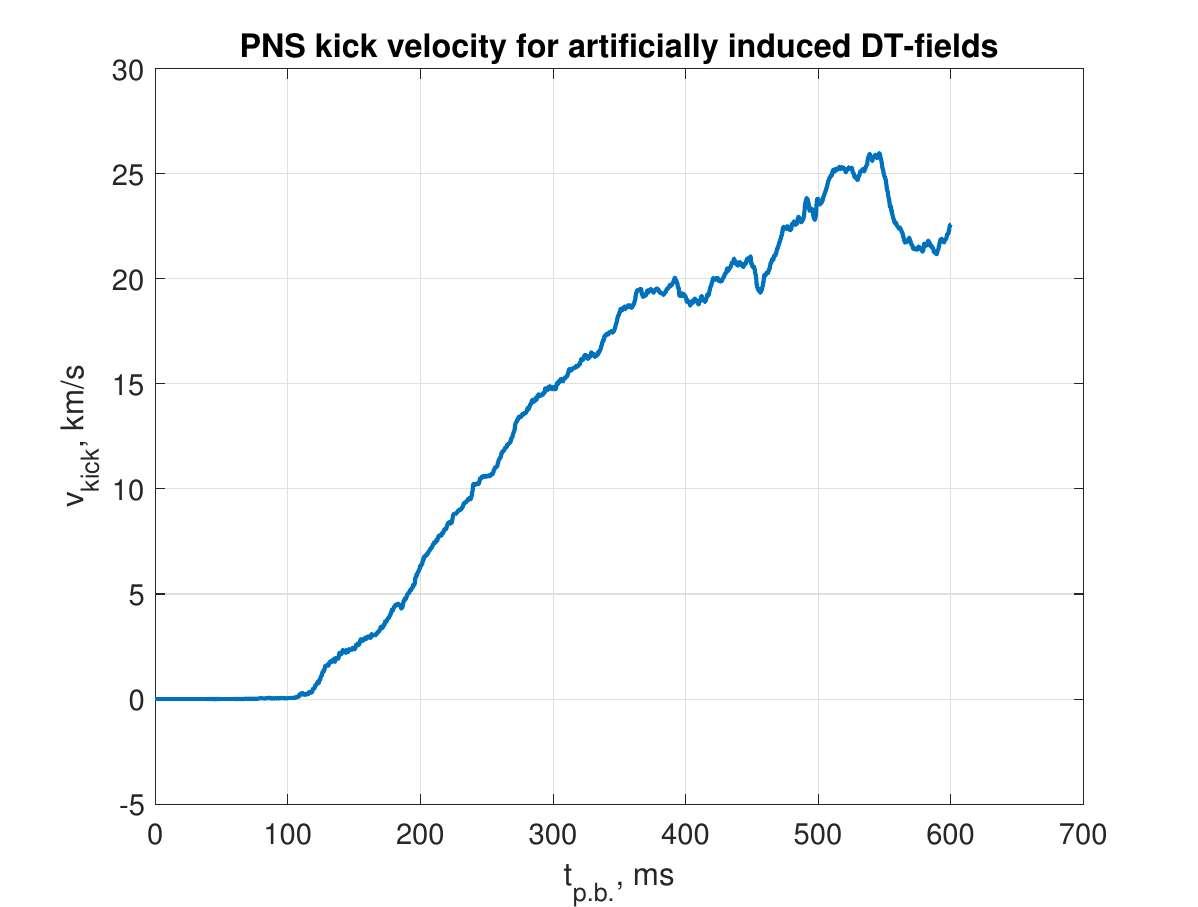}
		\includegraphics[width=2.2cm,height=4cm]{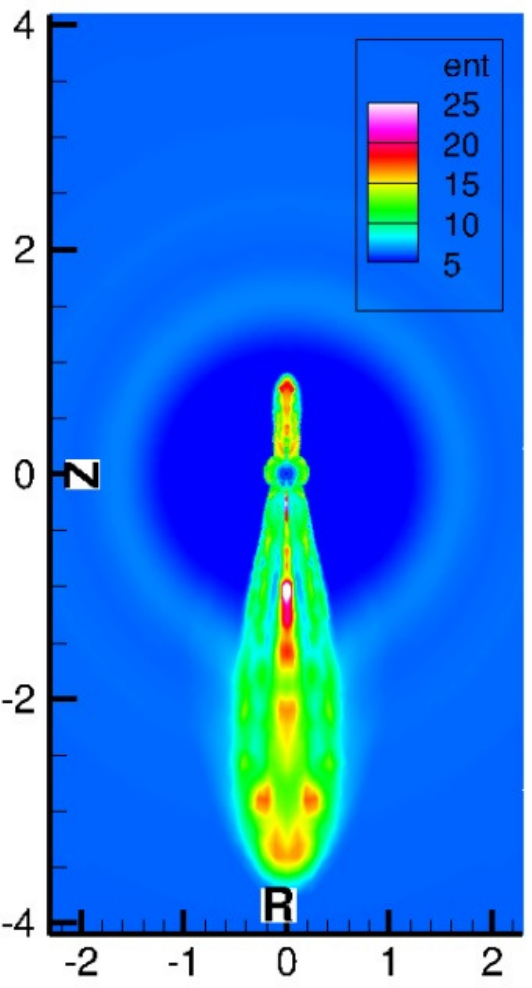}
		\caption{ The explosion properties of the test model with artificially induced poloidal and toroidal fields after the collapse. Upper panel -- explosion energy evolution, left lower panel -- kick velocity evolution, right lower panel -- entropy distribution ($k_B/m_u$) at $t_{p.b.} \approx 300$ ms. The axis scale on the right panel is in units of $1000$ km. }
		\label{fig:aux_dt_model}
	\end{figure}

It may be seen from Table \ref{tab:table_results_dt} and Fig. \ref{fig:b_phi_dt}, that the kick  has direction (opposite to jet asymmetry) in the side, where the toroidal magnetic field is increasing, in contrast to a simple model \cite{BKM1992}, where the poloidal field is always preserving its convex form. The concave form of the poloidal field acquired after the collapse leads to changing of the direction of the wrapped field and of the kick direction.    
To check this explanation we have conducted a construction of a simplified model with an artificially induced convex magnetic field after the core bounce. To set up the magnetic field distribution, we have used formula \eqref{magfield} with the parameters $B_{0,dip} = 4\cdot 10^{14}$ G, $B_{0,tor} = 2\cdot10^{16}$ G and $r_0 = 40$ km. The explosion energy and kick velocity for this model together with the entropy distribution at $t_{p.b.} \approx 300$ ms are presented in Fig. \ref{fig:aux_dt_model}. The equatorial asymmetry of the ejection formed here in the explosion has a direction opposite to the one produced by concave poloidal field, and the kick direction is in accordance with the model \cite{BKM1992}.

\section{Discussion and conclusions}

\begin{figure}[!htp]
		\centering
		\includegraphics[width=8.6cm,height=5.5cm]{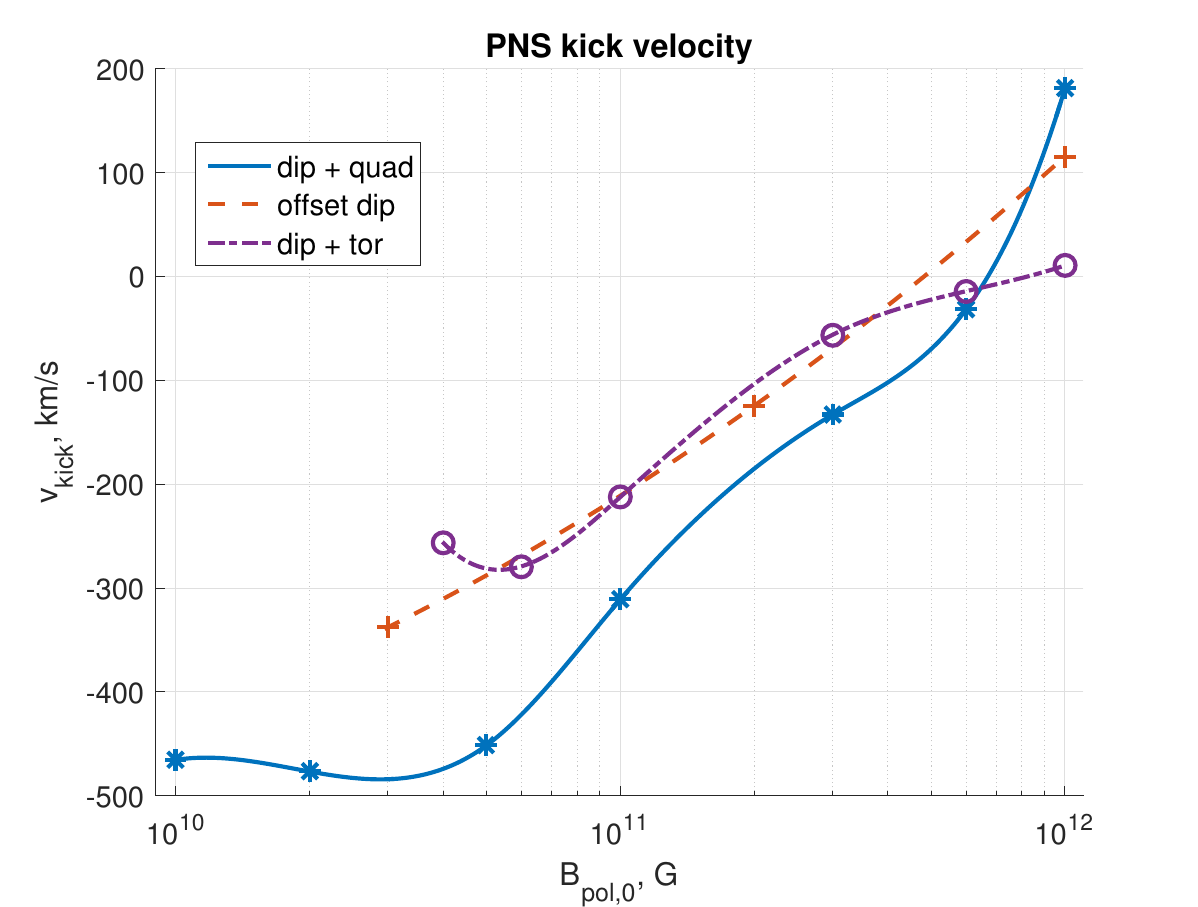}
		\caption{ The PNS kick dependence on the poloidal magnetic field for different simulated models at the final times of our simulations. The smoothed lines represent $DQ-$ (blue), $Do-$ (red dashed) and $DT-$ (violet dash-dotted) models. }
		\label{fig:kick_b}
	\end{figure}

In this work, we have examined an MHD-induced PNS kick generation in the framework of a MR supernova mechanism. Three possible options were studied. The first one is the superposition of quadrupolar and dipolar magnetic fields in the core of a progenitor star, the second case is with the initial offset dipolar field, and the third one is considering  the dipole field together with a symmetric  toroidal component before the core collapse. Due to the asymmetries in the outgoing jets, the MR mechanism can explain the gain of natal kicks of the protoneutron star up to $\sim 500$ km/s, formed during first few seconds after the core bounce. In the majority of simulated cases the kick velocity is  not saturated and could grow to larger values at later times. The kick velocity dependence on the magnetic field is plotted in Fig. \ref{fig:kick_b}, for three families of models, at the final times of simulations. 	
	
We have obtained for $DQ$ and $Do$ models that the highest asymmetry of the explosion occurs for lower magnetic fields with $B\,\, \sim few\,\, 10^{10}$ G (with $A_E \sim 0.7-0.8$), and decreases with increase of the poloidal magnetic field. The same stands for the kick velocity, which has a maximal absolute value ($v_{kick} \sim 480$ km/s for $DQ-2e10$ and $\sim 340$ km/s for $Do-3e10$). With increasing of the initial poloidal field strength, the anisotropy of explosion and the kick velocity decrease to zero, and at further rise change its direction. In $DQ$ models, both asymmetry value $A_E$ and the PNS velocity $v_{kick}$ reach almost zero values for the initial field  $B = 6\cdot 10^{11}$ G, and further it changes the sign and start to increase in the opposite direction. We see the kick saturation with time at an value of $\sim 180$ km/s for the model with the highest initial magnetic field $B_0 = 10^{12}$ G (Table \ref{tab:table_results_dq}, Fig.\ref{fig:expl_dq}). Note, that the model $DQ-1e12$ has a $\sim 15\%$ smaller explosion energy, than a test symmetric model (initial strong dipole field only, model $"D0-1e12"$ from Section \ref{sect51}) with the same initial magnetic energy, at the same post-bounce time. This result is in accordance with a recent study \cite{bugli2020}, where a set of core-collapse models with different magnetic multipoles was considered, and a decrease of the explosion energy with the multipole order was reported. The case of $Do$ models tends to be qualitatively similar to $DQ$ models with slightly smaller PNS velocities.  

In the set of $DT$ models, for the values of the poloidal magnetic field of the order of $\lesssim 10^{11}$ G the antisymmetric induced toroidal field formation together with a presence of the symmetric toroidal component in the PNS region leads to the formation of the PNS with kick in order of $\sim 200-300$ km/s. In these cases, the explosion develops stronger in the direction (above the equator in our calculations), where the winded toroidal field is summed up with the compressed component inside the first 10 km, and where the toroidal field passes through zero, leading to sharp gradient of the magnetic pressure. Such structure of the toroidal field appears due to the field wrapping in presence of the convex-concave poloidal magnetic field and a symmetric compressed component. With further increasing of the initial poloidal magnetic field, the toroidal component generated by differential rotation becomes dominant, and the system  $"$forgets$"$ about the symmetric part, leading to the explosions almost without  symmetry violation and the PNS kick. For the high magnetic field in the model $DT-1e12$, the explosion energy has $\sim 13\%$ higher values, than the one for the star with a purely dipolar field of the same strength (the test model $"D0-1e12"$ from Section \ref{sect51}), due to additional magnetic pressure gradient from the symmetric toroidal component.

Comparing the possible scenarios of the asymmetric jets formation in the context of MR supernovae, one can conclude, that the presence of asymmetric poloidal fields ($DQ$ and $Do$ models) in the collapsed core in the form of the multipoles composition produces  anisotropic explosions and PNS kick velocities more efficiently, than the model with an initial toroidal field ($DT$), especially for high initial magnetic fields. 
	
We have used two-dimensional axisymmetric MHD equations, where PNS kicks are aligned with the rotational axis. It would be interesting to consider a 3D setup for these problems, because some  \cite{kick2013,mueller2023} 3D numerical models are unable to reproduce the observed spin-kick alignment  \cite{smirnova1996,lai20012} (see, however, \cite{janka2022} and \cite{kick2023}). 

A simplified neutrino transfer treatment with a usage of the leakage scheme along the radial rays was implemented in our code. Such scheme allows to reproduce correctly the qualitative behaviour of neutrino physics in supernova environments. More accurate neutrino transfer simulations, using Monte-Carlo method (see e.g. \cite{kbk2018}) had been suggested for solving of this problem.

\begin{acknowledgments}
	
This work was supported by the Russian Science Foundation (RSF) grant 23-12-00198.
	
\end{acknowledgments}

\appendix
	
\subsection*{Appendix: Numerical Technique}

\begin{figure*}[!htp]
	\centering
	\includegraphics[width=5.5cm,height=4.9cm]{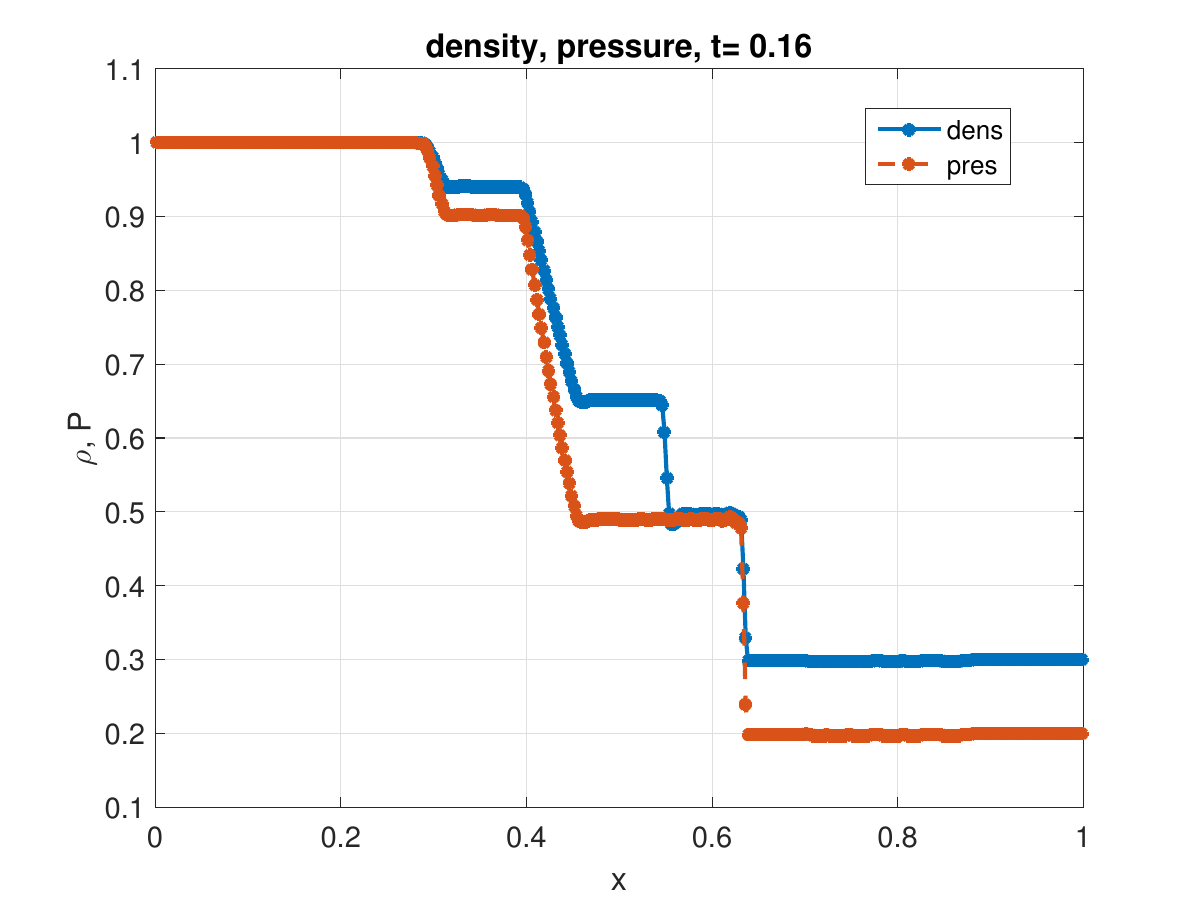}
	\includegraphics[width=5.5cm,height=4.9cm]{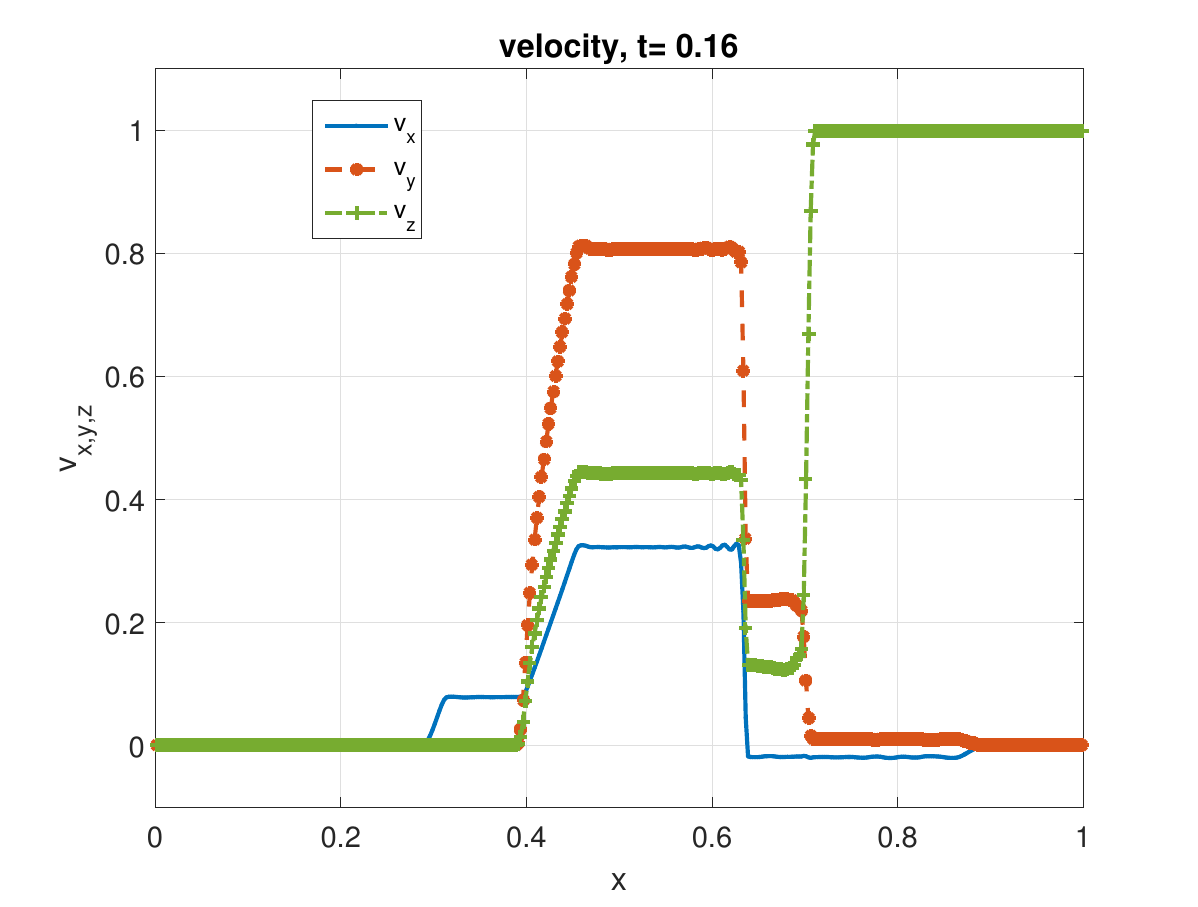}
	\includegraphics[width=5.5cm,height=4.9cm]{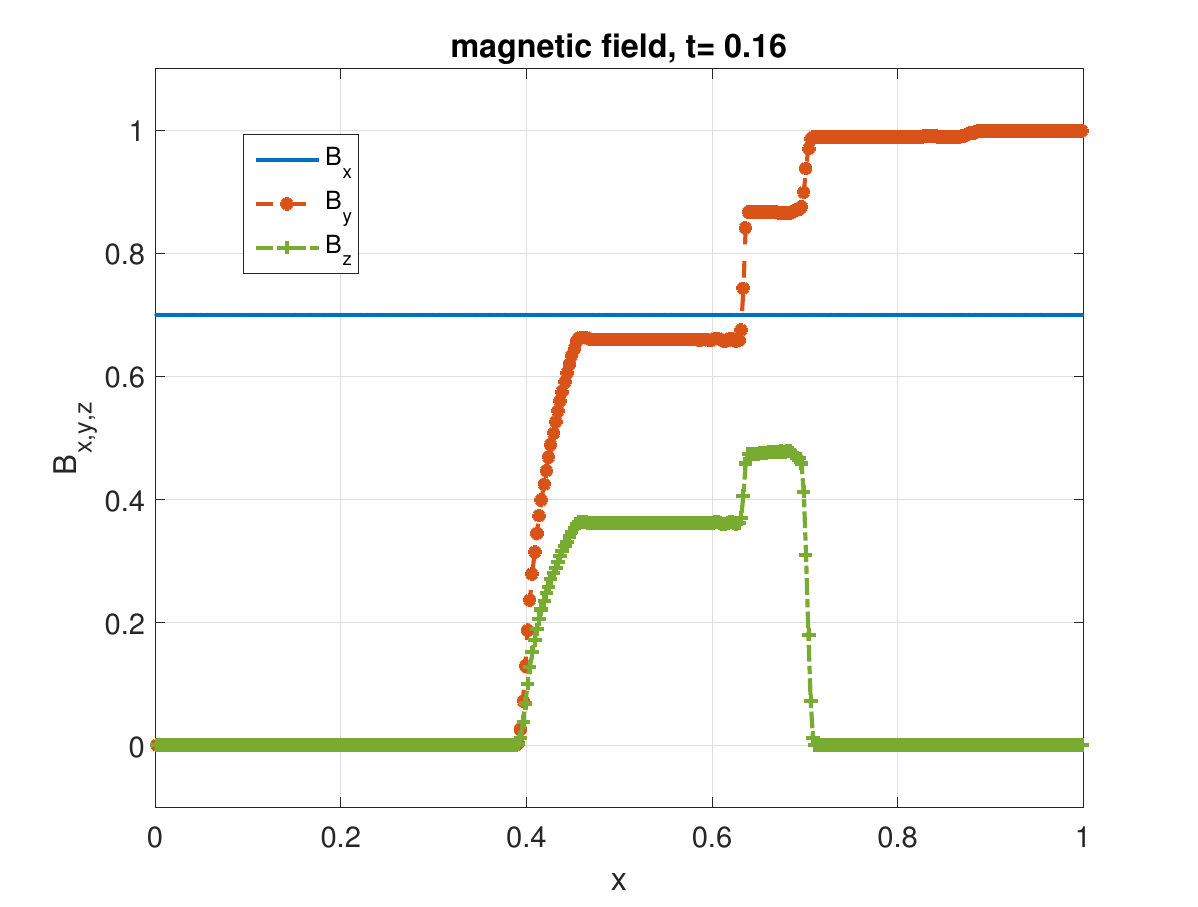}
	\caption{ Solution for the shock tube problem 4d from \cite{rd1995}: left panel -- density and pressure, central panel -- velocity field, right panel -- magnetic field. The EOS for ideal gas with $\gamma = 5/3$ is used. The time of the simulation is $t = 0.16$. The initial conditions are $(\rho, v_x,v_y,v_z, P, B_x,B_y,B_z) = (1.0,0.0,0.0,0.0,1.0,0.7,0.0,0.0)$ for $x<0.5$ and $(0.3,0.0,0.0,1.0,0.2,0.7,1.0,0.0)$ for $x>0.5$. The grid resolution is $N_x = 400$ cells. The results are in good accordance with \cite{rd1995}. 
	}
	\label{fig:mhd_tube}
\end{figure*}
	
We simulate the system \eqref{MHD} using our newly developed finite volume MHD solver \cite{ljm2024}, which utilizes the explicit Godunov-type methods for fluid equations by means of a piecewise linear method with a slope limiter or piecewise parabolic methods $PPM3$ and $PPM5$ \cite{mignone2014} for curvilinear coordinate systems. Near the coordinate singularities of the curvilinear geometries the Courant-Friedrichs-Lewy (CFL) stability condition can be very restrictive due to converging cell sizes. In order to mitigate this problem, the code adopts a $"$dendritic grid$"$ formalism as in $FORNAX$ code \cite{skinner2019}. In 2D spherical geometry it employs a grid rarefaction strategy in $\theta$-direction for the cells near the coordinate centre in the way, that the aspect ratio of the grid cell edges is in order of unity there. Hence, the CFL condition for the timestep basically depends only on the radial resolution and becomes less restrictive. Throughout this paper we use $HLLD$ approximate Riemann solver by \cite{miyoshi2005} in its less dissipative form proposed recently in \cite{imprHLLD}, as well as a fifth-order $PPM5$ reconstruction from \cite{mignone2014} for the primitive MHD variables together with explicit third-order $TVD$ Runge-Kutta time integrator $RK3$ \cite{so1988}. To approximately satisfy the magnetic field divergence constraint on the grid, we use a mixed hyperbolic-parabolic cleaning scheme by \cite{dedner2002}.

	\begin{figure}[!htp]
		\centering
		\includegraphics[width=4.2cm,height=3.6cm]{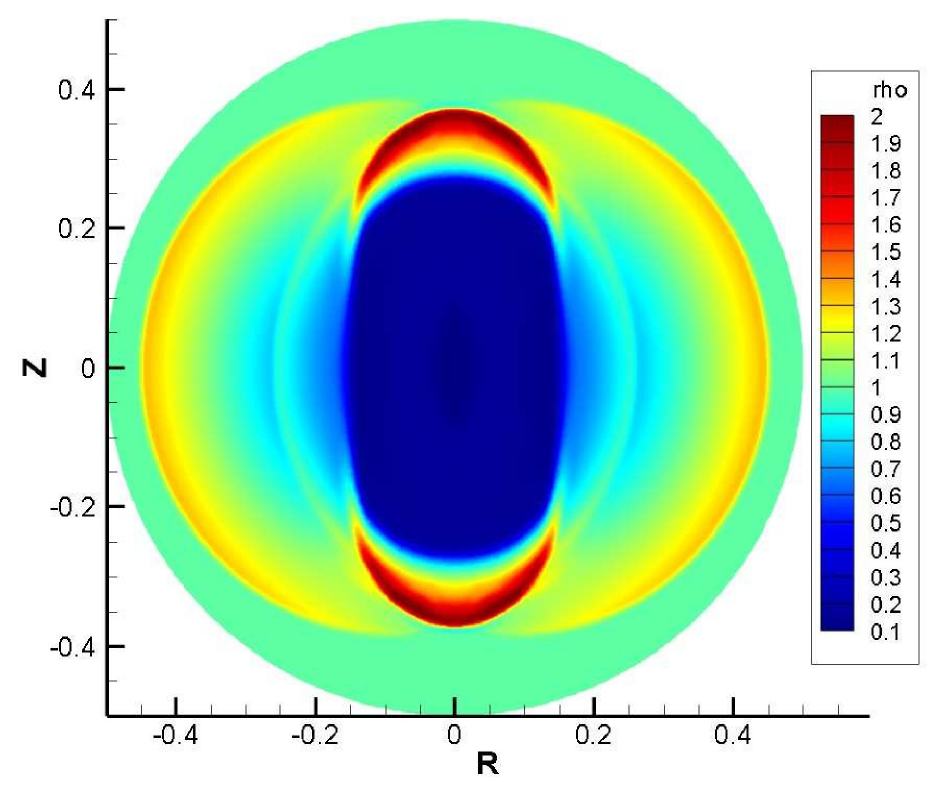}
		\includegraphics[width=4.2cm,height=3.6cm]{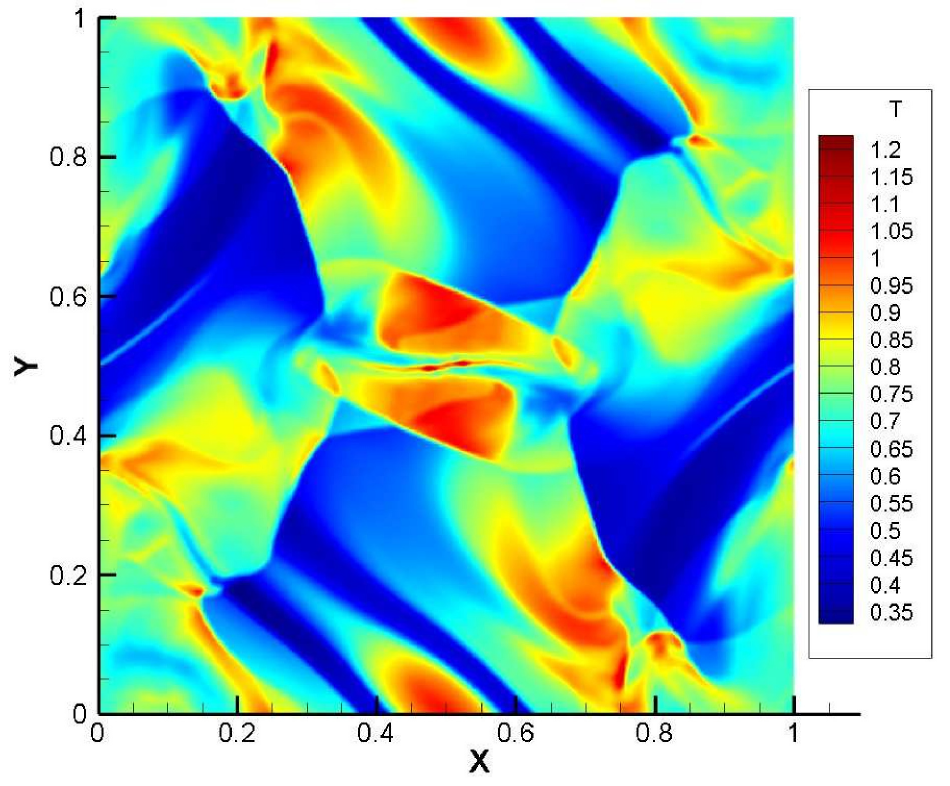}
		\caption{ Two-dimensional MHD tests: left panel -- density distribution in a 3D axisymmetric magnetized blast wave problem at time $t = 0.2$. The grid resolution is $N_r\times N_\theta = 128\times256$. We note, that due to the usage of the high-order volumetric reconstruction, we do not observe any spurious effects near the axis in this test. The initial conditions and reference solution can be found in, e.g., \cite{sg2009}; right panel -- temperature distribution in 2D Orszag-Tang vortex problem at time $t = 0.5$. The grid resolution is $N_x\times N_y = 256\times256$. The initial conditions and reference solution can be found in, e.g., \cite{matsumoto2019}.
		}
		\label{fig:mhd_2d}
	\end{figure}

To test the code, we have conducted several simulations of well-known MHD tests. The first one is a shock tube problem 4d from \cite{rd1995}, which involves all velocity and magnetic field components. The solution is presented in Fig. \ref{fig:mhd_tube}. To test the multidimensional ability of our code, we have calculated the magnetic blast wave test and Orszag-Tang vortex problem. The results are presented in Fig. \ref{fig:mhd_2d}.

\begin{figure}[!htp]
		\centering
		\includegraphics[width=4.2cm,height=3.5cm]{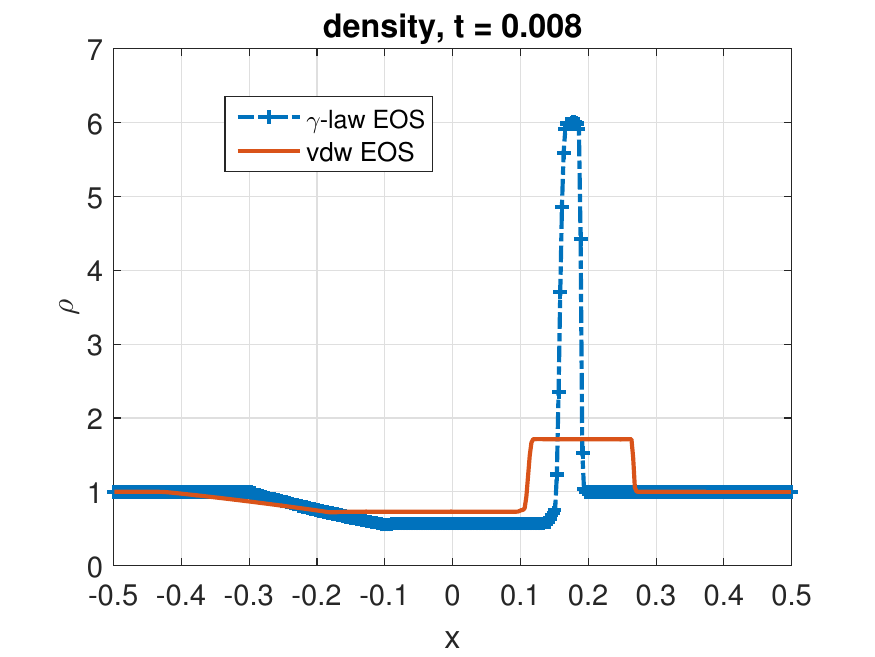}
		\includegraphics[width=4.2cm,height=3.5cm]{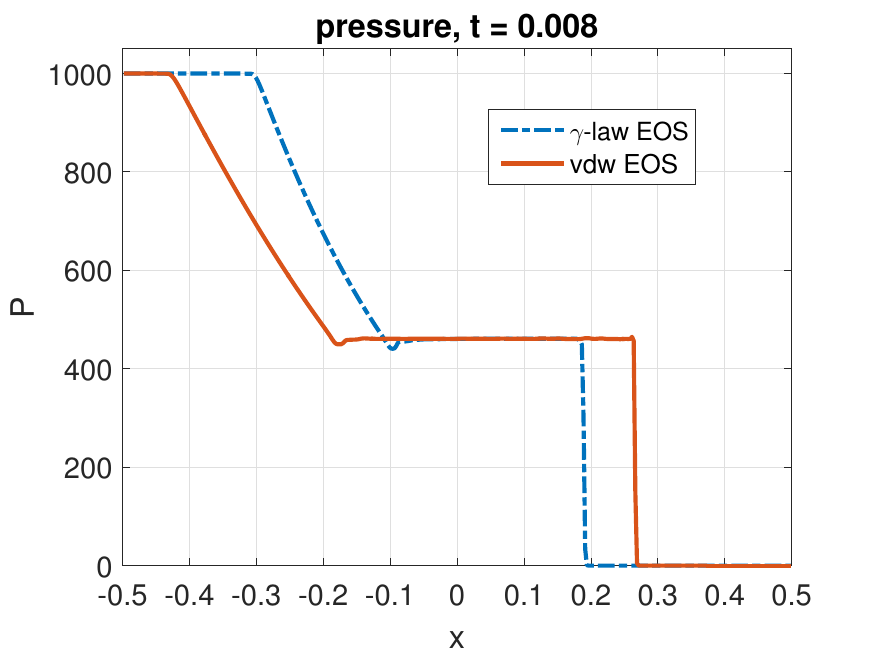}
		\caption{ Solution for the shock tube problem VDW3 from \cite{dc2016}: left panel -- density, right panel -- pressure. The time of the simulation is $t = 0.008$. The initial conditions are $(\rho, v_x,v_y,v_z, P) = (1.0, 0.0, 0.0, 0.0, 1000.0)$ for $x<0$ and $(1.0, 0.0, 0.0, 0.0, 0.01)$ for $x>0$. Blue lines correspond to ideal gas case with $\gamma = 7/5$, while red ones correspond to van-der-Waals EOS. The details about non-ideal EOS and reference solutions can be found in \cite{dc2016}. The grid resolution is $N_x = 400$ cells.
		}
		\label{fig:vdw_tube}
	\end{figure}

Finally, the supernova MHD code has to deal properly with a complex equation of state, written in a general form. To avoid the EOS call in the Riemann solver, we reconstruct internal energy $\rho e_{int}$ with other hydrodynamical variables and use its value for an energy flux calculation. To test this simple approach, we have solved the Riemann problem VDW3 from \cite{dc2016} with van-der-Waals EOS. It involves a strong shock, a contact discontinuity and a rarefaction wave, which have different properties in comparison to ideal $\gamma$-law EOS case. The results are given in Fig.\ref{fig:vdw_tube}. The matching of the presented tests to the published ones concludes the validity of the results, obtained with our MHD code.

	\bibliography{MR_NS_kicks}

\end{document}